\documentclass[%
 aip,
 amsmath,amssymb,
 reprint,%
]{revtex4-1}

\usepackage{graphicx}
\usepackage{dcolumn}
\usepackage{chemfig}
\usepackage{amsmath}
\usepackage{amssymb}

\usepackage[utf8]{inputenc}
\usepackage[T1]{fontenc}
\usepackage{mathptmx}
\usepackage{etoolbox}
\usepackage{float}

\newcommand{\Sec}[1]{Sec.\,\ref{#1}}
\newcommand{\Eq}[1]{Eq.\,(\ref{#1})}
\newcommand{\Fig}[1]{Fig.\,\ref{#1}}

\newcommand{\RNum}[1]{\uppercase\expandafter{\romannumeral #1\relax}}
\usepackage{tikz}

\makeatletter
\def\@email#1#2{%
 \endgroup
 \patchcmd{\titleblock@produce}
  {\frontmatter@RRAPformat}
  {\frontmatter@RRAPformat{\produce@RRAP{*#1\href{mailto:#2}{#2}}}\frontmatter@RRAPformat}
  {}{}
}%
\makeatother
\begin{document}

\preprint{AIP/123-QED}

\title{Probing voltage-induced chemical reactions and anharmonicity with a confined vacuum light field}
\author{Yaling Ke}
\email{yaling.ke@phys.chem.ethz.ch}
\affiliation{ 
Department of Chemistry and Applied Biosciences, ETH Zürich, 8093 Zürich, Switzerland
}%

\begin{abstract}
In this work, we present a proof-of-concept investigation of non-equilibrium chemical reaction dynamics at a molecule-electrode interface, driven out of equilibrium by an applied votage bias and mediated by a confined, enhanced vacuum electromagnetic field inside an optical cavity. The coupled electron-vibration-photon system, together with the electrodes and a dissipative environment, is described within an open quantum system framework and solved using a numerically exact quantum dynamical approach.  The reaction coordinate is modeled with a Morse potential, enabling explicit treatment of molecular anharmonicity and bond-breaking behavior. By varying the cavity frequency across the infrared regime to cover typical nuclear vibrational energies, we observe multiple resonant rate suppression features that emerge whenever the cavity mode is brought into resonance with a dipole-allowed vibrational transition along the anharmonic ladder up to the dissociation threshold.  These findings open the door to extending polaritonic chemistry into genuinely nonequilibrium scenarios relevant to molecule-electrode interfaces.  Moreover, building on these results, we further propose a multi-mode vibrational strong coupling strategy in which several cavity modes are individually matched to distinct vibrational transitions. This engineered multi-resonant cavity induces a stepwise vibrational ladder descending process that efficiently drains vibrational excited energy. The resulting cavity-assisted cooling suggests a potential route toward mitigating voltage-induced bond rupture and the long-standing instability issues of molecular junctions operating under high bias.
\end{abstract}

\maketitle

\section{Introduction}
In pursuit of active and selective control over molecular properties and chemical reactivity, strong coupling to the vacuum electromagnetic field has emerged as a promising strategy, offering a non-invasive and highly tunable platform for manipulating molecular dynamics.\cite{Ebbesen_2016_ACR_p2403,Kasprzak_Nat_2006_p409,AberraGuebrou_2012_PRL_p66401,Coles_Nat.Mater._2014_p712,Orgiu_2015_NM_p1123,Hagenmueller_2017_PRL_p223601,Hsu_2017_JPCL_p2357,Baieva_2017_AP_p28,Mony_2021_AFM_p2010737,Hayashi_2024_JCP_p181101,Sandik_2025_NM_p344}Among the many intriguing phenomena revealed in the vibrational strong-coupling regime,\cite{Dunkelberger_2016_NC_p13504,Muallem_2016_JPCL_p2002,Xiang_2020_S_p665,Hirai_2021_CS_p11986a,Joseph_2021_ACIE_p19665,Fukushima_2022_JACS_p12177,Wright_2023_JACS_p5982,Hou_2024_AP_p1303,Menghrajani_2019_AOM_p1900403,Damari_2019_NC_p3248} where molecular vibrations hybridize with infrared photons to form vibrational polaritons, the most celebrated advances in the burgeoning field of polaritonic chemistry have been the observations of resonant modifications to chemical reaction rates inside optical microcavities.\cite{Thomas_2016_ACE_p11634,Vergauwe_2019_ACIE_p15324,Lather_2019_ACIE_p10635,Thomas_2019_S_p615,Hiura_2019__p,Hirai_2020_ACE_p5370,Pang_2020_ACIE_p10436,Sau_2021_ACIE_p5712,Ahn_Sci_2023_p1165,Patrahau_Angew.Chem.Int.Ed._2024_p202401368,Mahato_2025_ACIE_p202424247} However, despite rapid growth over the past decade, progress has been hampered by the lack of a clear mechanistic understanding of the underlying mechanisms and by a number of controversial  experimental results,\cite{Imperatore_2021_JCP_p191103,Wiesehan_2021_JCP_p241103,Thomas_2024_JPCL_p1708,Chen_2024_N_p2591, Muller_2024_ACIE_p202410770, Muller_2025_ACIE_p202509391,Jin_2025_JACS_p38320} which have collectively slowed the field’s momentum.
 
Despite significant challenges, recent efforts--spanning both theoretical and experimental fronts--have been devoted to pushing past this bottleneck.  On the theoretical side, fully quantum dynamical simulations have successfully captured the sharp resonant peaks in the cavity-frequency-dependent rate modification profile in the few-molecule limit.\cite{Lindoy_2023_NC_p2733,Lindoy_2024_N_p2617,Ying_2023_JCP_p84104,Hu_2023_JPCL_p11208,Ying_2024_CM_p110,Ke_J.Chem.Phys._2024_p224704,Ke_2024_JCP_p54104,Ke_J.Chem.Phys._2025_p64702,Ke_2025_JCP_p54109,Ke_2025_JCP_p164703} Experimentally, a variety of novel cavity architectures have been developed that extend beyond traditional microfluidic Fabry-P\'erot geometries. For example, to address difficulties associated with  sampling and rate measurements, a continuous-flow Fabry-P\'erot cavity integrated with real-time spectroscopic monitoring has been constructed.\cite{Yamada_2022_JPCB_p4689,Lian_2025_AP_p3557} To circumvent mirror-alignment issues in Fabry-P\'erot cavities, nonlocal metasurfaces formed by arrays of plasmonic particles that are capable of supporting highly tunable and enhanced surface lattice resonance modes have also been employed, demonstrating vibrational strong coupling and rate modifications.\cite{Verdelli_2024_ACIE_p202409528} Yet, reactive dynamics studied to date inside confined radiation fields have largely been restricted to thermally activated reactions near equilibrium. 

In this work, we carry out a proof-of-concept theoretical exploration of polaritonic chemistry under voltage-driven nonequilibrium conditions, with the aim of  extending the scope of cavity-modified reactions to the regimes that are inaccessible under standard thermal environments. Although the proposed setup is hypothetical, it may become experimentally feasible in the future through the compatible integration of a molecular junction—where a single molecule bridges either metallic leads or thin layers of graphene electrodes\cite{Nitzan_Science_2003_p1384,Thoss_J.Chem.Phys._2018_p30901}—with a confined and strongly enhanced electromagnetic field.\cite{Climent_2019_ACIE_p8698,Nagarajan_2020_AN_p10219,Spencer_2025_LSA_p69} We describe the entire architecture as an open quantum system, in which the molecular electronic and vibrational degrees of freedom (DoFs) together with discretized photonic modes constitute the central system of interest. The electrodes, the cavity loss into the far-field electromagnetic continuum, and the substantial dissipative environment associated with the liquid phase and surface phonons are modeled as fermionic and bosonic reservoirs that exert essential influence over the system dynamics.

Our study focuses specifically on the bond rupture dynamics along a stretching reaction coordinate evolving on a Morse potential, which naturally incorporates molecular anharmonicity and dissociative behavior. The full quantum-mechanical dynamics of the composite electron–vibration–photon system are computed using the numerically exact hierarchical equations of motion (HEOM) method (see the review in Ref.~\onlinecite{Tanimura_2020_JCP_p20901} and references therein) implemented with a compact and efficient tensor network state representation.\cite{Ke_2023_JCP_p211102}  We find that tuning an infrared cavity mode into resonance with dipole-allowed nearest-neighbor or overtone vibrational transitions leads to sharp resonant suppression of ultrafast reaction rates, suggesting that confined vacuum optical fields may serve as sensitive probes of reactive molecular anharmonicity. Moreover, when multiple cavity modes are introduced such that each is resonant with a distinct vibrational transition within the vibrational manifold, a cascading vibrational ladder descending pathway opens up. This engineered radiative cooling channel efficiently drains vibrational energy from highly excited states and may provide a promising strategy for stabilizing current-carrying molecular electronic junctions under high voltage biases.

The remainder of the paper is organized as follows. In \Sec{sec:theory}, we introduce the microscopic model, theoretical approaches, and simulation details used to study voltage-induced nonequlibrium chemical reactions inside an optical microcavity. \Sec{sec:results} presents the bond-dissociation dynamics in the absence and presence of the cavity field, under both single-mode or multi-mode strong coupling conditions. Finally,  \Sec{sec:conclusion} summarizes our key findings and outlines future directions for advancing this line of research.

\begin{figure*}
\centering
 \begin{minipage}[c]{0.53\textwidth}
    \raggedright a) 
    \includegraphics[width=\textwidth]{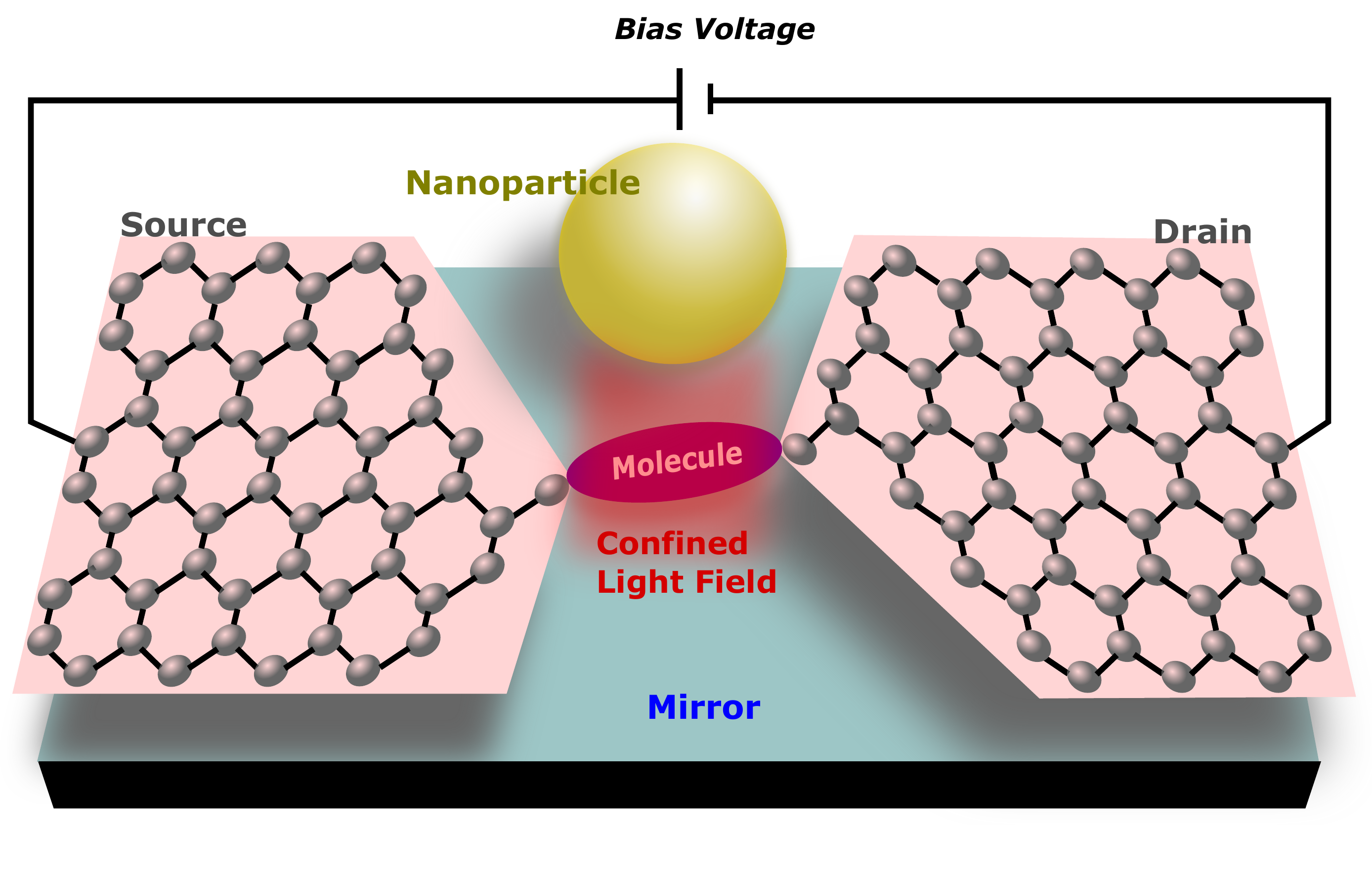}
 \end{minipage}
 \begin{minipage}[c]{0.45\textwidth}
    \raggedright  b) 
    \includegraphics[width=\textwidth]{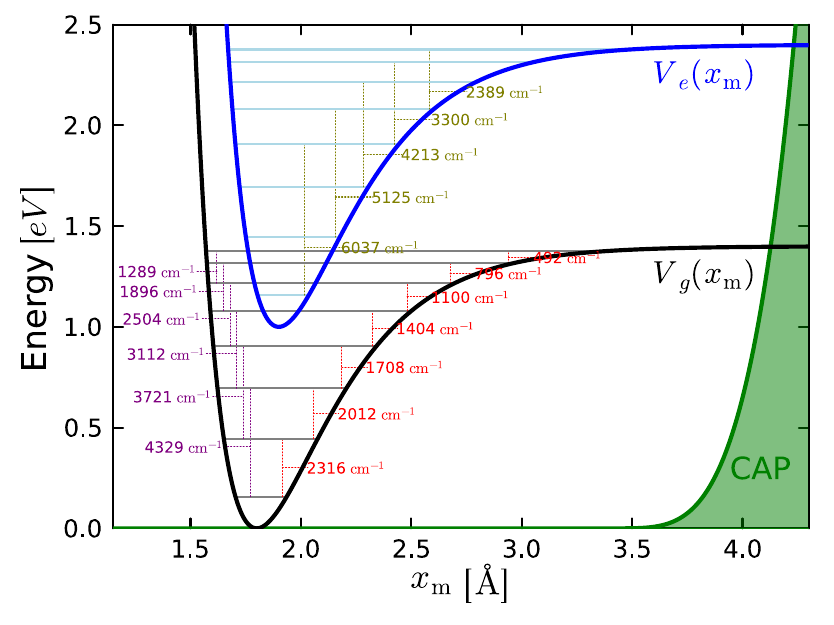}
  \end{minipage}
   \begin{minipage}[c]{0.7\textwidth}
    \raggedright  c) 
    \includegraphics[width=\textwidth]{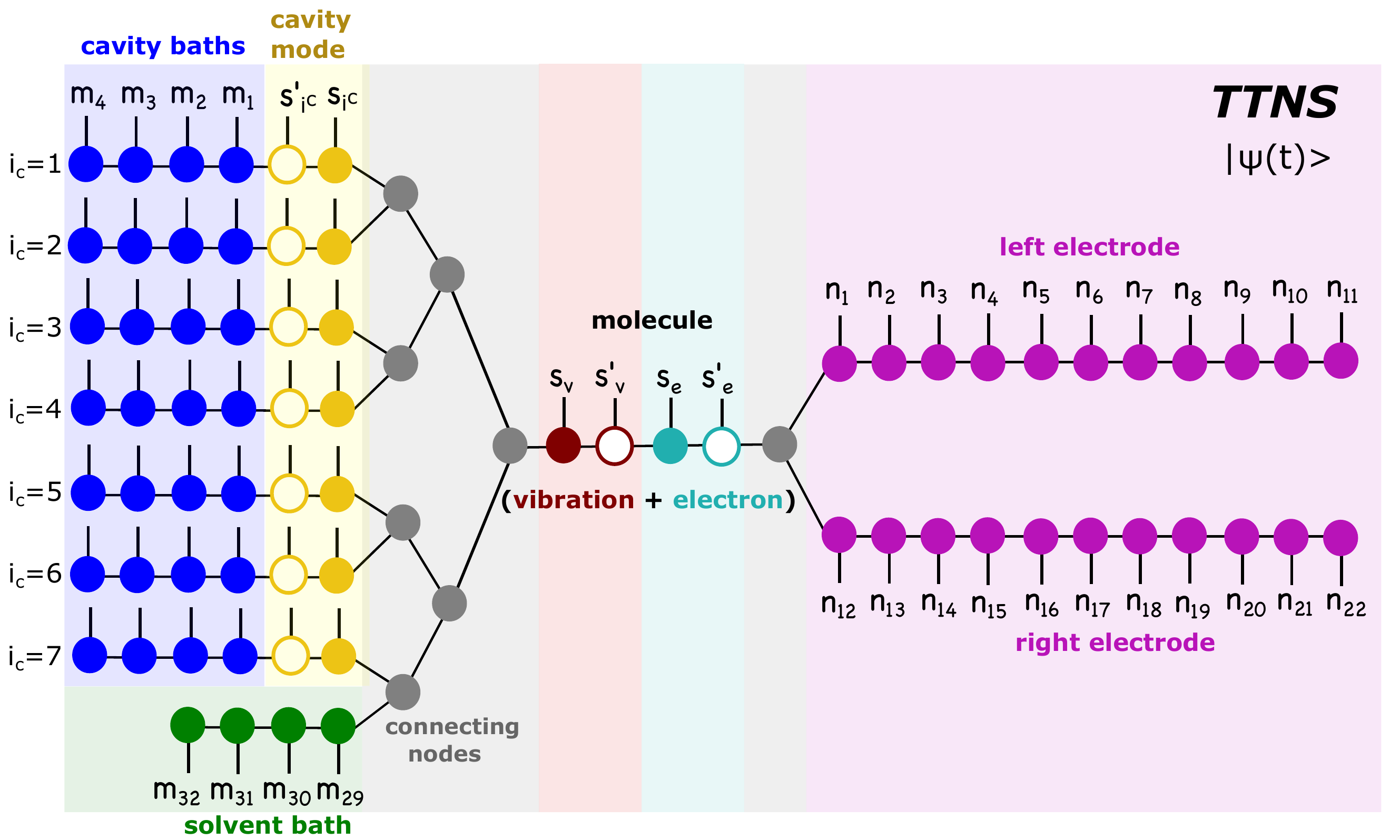}
  \end{minipage}
\caption{a) Schematic illustration of a single molecule bound to two electrodes under an applied bias voltage $\Phi$, and confined within an enhanced optical field provided by a nanoparticle-on-mirror cavity structure.  b) Morse potential energy surfaces of the neutral and charged electronic states, characterized by $D_{\rm g}=D_{\rm e}=1.4$~eV, $a_{\rm g}=a_{\rm e}=3\text{\r{A}}$, $x_{\rm g}=1.8\text{\r{A}}$, and $x_{\rm m}^{e.0}=1.9\text{\r{A}}$. The charging energy is $E=1$~eV. The green shaded region shows the absolute value of the CAP $|V_{\rm CAP}(x_{\rm m})|$. Thin horizontal lines indicate the bound vibrational eigenstate energies of the isolated molecule below the dissociation barrier. The dotted red, purple, and olive-green vertical lines denote the nearest-neighbor, two-quantum, three-quantum vibrational transitions, respectively, with the corresponding energy gaps labeled on the side. We note that the coupling to the cavity and its associated bath, as well as the electrodes and solvent environment might broaden these levels and can shift the transition energies.  c) Illustration of the tree tensor network state decomposition of the high-dimensional coefficient tensor $C_{\bf{s}, \bf{s}'}^{ \bf{n}, \bf{m}}(t)$ appearing in the extended wavefunction $|\Psi(t)\rangle$ [defined in \Eq{ExtendedWaveFunction}] for the full molecule-cavity-environment composite system.
} \label{fig1:schematic}
\end{figure*}
\section{\label{sec:theory}Theory}
\subsection{Model and Method}
To investigate voltage-driven non-equilibrium reaction dynamics in the presence of a confined vacuum electromagnetic field, we construct a hypothetical model system that is sufficiently complex to capture the essential interplay among molecular, photonic, and environmental degrees of freedom (DoFs), as schematically illustrated in \Fig{fig1:schematic}~a), yet simple enough to allow a fully quantum-mechanical and numerically exact treatment of all DoFs on equal footing.  The model is formulated within an open quantum system framework, with the total Hamiltonian written as 
\begin{equation}
    H = H_{\rm S} + H_{\rm B}. 
\end{equation}

The central system consists of a single molecule coupled to discrete photonic modes, which may represent standing waves confined between the parallel mirrors of a Fabry-P\'erot cavity\cite{GarciaVidal_2021_S_p336} or localized surface plasmons supported on the metasurfaces of metallic nanoparticles or arrays.\cite{Barnes_2003_N_p824,Toermae_2014_RPP_p13901} Notably, vibrational strong coupling has recently been demonstrated for a single or a few molecules placed in a nanoparticle-on-mirror cavity geometry.\cite{Chikkaraddy_2016_N_p127} The light-matter interaction is described by the Pauli-Fierz quantum electrodynamics Hamiltonian in the dipole gauge under the long-wavelength approximation.\cite{Flick_2017_PotNAoS_p3026,Rokaj_2018_JPBAMOP_p34005,Mandal_2023_CR_p9786,Lindoy_2023_NC_p2733}  This formulation correctly captures the interaction between a molecule and the quantized transverse electromagnetic field in the non-relativistic limit (hereafter, we set $\hbar=e=k_{\rm B}=1$): 
\begin{equation}
\label{SystemHamiltonian}
\begin{split}
    H_{\rm S} = & \frac{p_{\rm m}^2}{2M_{\rm S}}+V_{\rm g}(x_{\rm m}) + \left(V_{\rm e}(x_{\rm m})-V_{\rm g}(x_{\rm m})\right)d^{+}d^{-} \\ 
 &   +\sum_{i_{\mathrm{c}}} \frac{p_{i_{\mathrm{c}}}^2}{2} +  \frac{1}{2} \omega_{i_{\mathrm{c}}}^2\left(x_{i_{\mathrm{c}}} - \sqrt{\frac{2}{\omega_{i_{\mathrm{c}}}}} \eta_{i_{\mathrm{c}}} \vec{u}(x_{\mathrm{m}})\cdot \vec{e} \right)^2.  
\end{split}
\end{equation}

The molecule is characterized by a single spinless electronic orbital and a reactive intramolecular vibrational mode. The reaction coordinate, conjugate momentum, and the reduced mass are denoted by $x_{\mathrm{m}}$, $p_{\mathrm{m}}$, and $M_{\rm S}$, respectively. Electron creation and annihilation on the molecular orbital are represented by the operators $d^{+}$ and $d^{-}$. The neutral-state potential energy surface (PES) along the reaction coordinate is modeled by a Morse potential,
\begin{equation}
    V_{\rm g}(x_{\rm m}) = D_{\rm g}\left(1-e^{-(x_{\rm m}-x_{\rm m}^{\rm g,0})/a_{\rm g}})\right)^2,
\end{equation}
with dissociation energy $D_{\rm g}$, width parameter $a_{\rm g}$, and equilibrium position $x_{\rm m}^{\rm g,0}$, chosen to describe chemical bond breaking.\cite{Ke_J.Chem.Phys._2021_p234702} The energy at the neutral equilibrium geometry is taken as the zero reference. When the orbital is occupied, the charged-state PES also follows a Morse form:
\begin{equation}
      V_{\rm e}(x_{\rm m}) = D_{\rm e}\left(1-e^{-(x_{\rm m}-x_{\rm m}^{\rm e,0})/a_{\rm e}})\right)^2+E,
\end{equation}
with charging energy $E$ and parameters: $D_{\rm e}$, $a_{\rm e}$, and $x_{\rm m}^{e,0}$. The charging energy $E$ is constant and may in principle be adjusted using a gate voltage. Extension to multiple electronic orbitals and vibrational modes is straightforward but omitted here for clarity. Representative PESs of $V_{\rm g}(x_{\rm m})$ and $V_{\rm e}(x_{\rm m})$ are shown in \Fig{fig1:schematic}~b).

Each quantized cavity modes is described as a harmonic oscillator with frequency $\omega_{i_{\rm c}}$, coordinate $x_{i_{\rm c}}$, momentum $p_{i_{\rm c}}$, and unit mass. The cavity frequency can be varied by adjusting mirror separation of confined optical setups or the size of plasmonic nanoparticles. Light-matter interaction displaces each oscillator proportionally to the projection of the molecular electronic dipole moment $\vec{u}(x_{\rm m})$ onto the light polarization direction $\vec{e}$, scaled by the coupling strength $\eta_{i_{\rm c}}$. Throughout this work, we assume that the molecular dipole is oriented to align with the cavity polarization to maximize the molecule-cavity coupling. To concentrate on vibrational strong coupling effects and their influence on reaction dynamics, the cavity frequency is restricted to the infrared (IR) spectrum range, covering typical molecular vibrations. Due to the energetic mismatch, we assume that electronic transitions between the neutral and charged states can not be driven by cavity photons and set $u_{eg}=0$.  Besides, $u_{gg}(x_{\rm m})=u_{ee}(x_{\rm m})$ and they remain dependent on the nuclear configuration $x_{\rm m}$. 

The molecule is connected to two macroscopic electrodes (metallic, semiconducting, or graphene-based). For instance, as illustrated schematically in  \Fig{fig1:schematic}~a), the molecule may be anchored between two thin graphene layers. A voltage bias $\Phi$, representing the free energy input that drives the electrochemical reaction away from equilibrium, is assumed to drop symmetrically across the junction on both electrodes, yielding chemical potentials $\mu_L=-\mu_R=\Phi/2$. Each electrode is modeled as a fermionic reservoir comprising numerous non-interacting electrons, forming continuous electronic states: 
\begin{equation}
\label{Ebath}
    H_{\rm ebath} = \sum_{\alpha \in\{L,R\}}\sum_k \epsilon_{\alpha k}c^{+}_{\alpha k}c^-_{\alpha k}
    + v_{\alpha k}c_{\alpha k}^{+}d^{-}+ v^*_{\alpha k}d^{+}c^{-}_{\alpha k}.
\end{equation}
Here, $\alpha$ specifies the left or right electrode, $c_{\alpha k}^{+}/c_{\alpha k}^{-}$ creates/annihilates an electron with the energy $\epsilon_{\alpha k}$ in the $k$th state of electrode $\alpha$.  Electron exchanges between the molecule and electrodes, and the transfer rate is governed by the coupling coefficients $v_{\alpha k}$, which are assumed to be independent of the molecular nuclear and photonic DoFs. However, the energy influx carried by an incoming electron can be funneled into the reactive vibrational mode through vibronic coupling, thereby activating or accelerating otherwise inaccessible or slow thermal reactions. 

Energy dissipation from the central system occurs through interactions with the solvent, electrode phonons, and cavity loss. These processes are modeled by introducing bosonic baths of infinite harmonic oscillators. The bath responsible for molecular vibrational relaxation is
\begin{equation}
    H_{\rm mbath} = \sum\limits_{k}\frac{\mathrm{P}^2_{{\rm m} k}}{2} + \frac{1}{2}\omega^2_{{\rm m} k}\left( \mathrm{Q}_{{\rm m} k} - \frac{g_{{\rm  m} k}h_{\rm m}(x_{\rm m})}{\omega_{{\rm, m} k}^2}\right)^2,
\end{equation}
where we assume the coupling operator in the system subspace to be $h_{\rm m}(x_{\rm m})= x_{\rm m}-x_{\rm m}^{g,0}$, corresponding to displacement from the neutral equilibrium position. Cavity losses are modeled similarly, by coupling each cavity mode to a cavity bath:
\begin{equation}
    H_{\rm cbath} = \sum_{i_{\mathrm{c}}}\sum\limits_{k}\frac{\mathrm{P}^2_{i_{\mathrm{c}} k}}{2} + \frac{1}{2}\omega^2_{i_{\mathrm{c}} k}\left( \mathrm{Q}_{i_{\mathrm{c}} k} - \frac{g^{}_{i_{\mathrm{c}} k}h^{}_{i_{\mathrm{c}}}(x^{}_{i_{\mathrm{c}}})}{\omega_{i_{\mathrm{c}} k}^2}\right)^2.
\end{equation}
Here, the $k$th oscillator in bath $\theta$, associated with the coupling to either the molecule $(\theta={\rm m})$) or a cavity mode $(\theta=i_{\mathrm{c}})$ is characterized by coordinate $Q_{\theta k}$, momentum $P_{\theta k}$, and frequency $\omega_{\theta k}$, and its equilibrium is shifted proportionally to the coupling strength $g_{\theta k}$, along with the operator $h_{i_{\mathrm{c}}}(x^{}_{i_{\mathrm{c}}})=x^{}_{i_{\mathrm{c}}}$.

The statistical information of the hybrid fermionic-bosonic Gaussian environment, $H_{\rm B}=H_{\rm ebath}+H_{\rm mbath}+H_{\rm cbath}$, and the influence on system dynamics, are fully encoded in their corresponding time correlation functions. 

For the electrodes, the correlation function is given by
\begin{equation}
     C_{\alpha}^{\sigma}(t) = \frac{1}{2\pi} \int_{-\infty}^{\infty}  e^{i\sigma \epsilon t}\Gamma_{\alpha}(\epsilon)f_{\alpha}^{\sigma}(\epsilon)\mathrm{d}\epsilon ,  
\end{equation}
where $\sigma=\pm$ stands for hole and electron, respectively, and we use the shorthand $\bar{\sigma} = -\sigma$. The Fermi function $f_{\alpha}^{\sigma}(\epsilon) = \frac{1}{1+e^{\sigma\beta_{\alpha}(\epsilon-\mu_{\alpha})}}$ characterizes the electronic occupation distribution in electrode $\alpha$ at inverse temperature $\beta_{\alpha}=1/(k_{\rm B}T_{\alpha})$. The spectral density function, which is also commonly referred to as the level-width function or hybridization function, is defined as
\begin{equation}
    \Gamma_{\alpha}(\epsilon) = 2\pi \sum_k |v_{\alpha k}|^2\delta(\epsilon-\epsilon_{\alpha k}).
\end{equation} 
For the bosonic environment, the correlation function takes the form
\begin{equation}
     G_{\theta}(t) = \frac{1}{2\pi} \int_{-\infty}^{\infty} e^{-i\omega t}J_{\theta}(\omega)f_{\theta}(\omega)  \mathrm{d}\omega 
\end{equation}
with the Bose-Einstein distribution function $f_{\theta}(\omega)=\frac{1}{e^{\beta_{\theta} \omega}-1}$ and the bosonic spectral density 
\begin{equation}
    J_{\theta}(\omega) = 2\pi \sum_k\frac{g_{\theta k}^2}{2\omega_{\theta k}}\delta(\omega-\omega_{\theta k}).
\end{equation}

Both the fermionic and bosonic bath correlation functions can be decomposed into sums of exponential terms:
\begin{subequations}
\label{BCF_exp}
\begin{equation}
\label{FermiBCF_exp}
     C_{\alpha}^{\sigma}(t) \cong \sum_{p=0}^{P^{\rm F}\rightarrow \infty} (\lambda^{\rm F}_{\alpha})^2\eta^{\rm F}_{\alpha p \sigma}e^{-\gamma^{\rm F}_{\alpha p \sigma}t},  
\end{equation}
\begin{equation}
\label{BosonBCF_exp}
     G_{\theta}(t) \cong \sum_{p=0}^{P^{\rm B} \rightarrow \infty} (\lambda^{\rm B}_{\theta})^2 \eta^{\rm B}_{\theta p}e^{-\gamma^{\rm B}_{\theta p}t}.
\end{equation}   
\end{subequations}

This exponential decomposition forms the basis of the HEOM method. Specifically,
by interpreting each exponential term in Eqs.~\eqref{FermiBCF_exp} and \eqref{BosonBCF_exp} as an effective pseudomode, one can construct a Schr\"odinger-like equation\cite{Borrelli_2019_JCP_p234102,Borrelli_2021_WCMS_p1539,Ke_2022_JCP_p194102} 
\begin{equation}
\label{SchroedingerEquation}
    i\frac{d|\Psi(t)\rangle}{dt}=\mathcal{H}|\Psi(t)\rangle
\end{equation}
for an extended wavefunction defined as
\begin{equation}
\label{ExtendedWaveFunction}
    |\Psi(t)\rangle = \sum_{\bf{n}}\sum_{\bf{m}}\sum_{\bf{s}, \bf{s}'}C_{\bf{s}, \bf{s}'}^{ \bf{n}, \bf{m}}(t) |\bf{s}, \bf{s}'\rangle \otimes |\bf{n}\rangle \otimes |\bf{m}\rangle,
\end{equation}
where the auxiliary indices $\mathbf{n}$ and $\mathbf{m}$ encode the excitations of the fermionic and bosonic pseudomodes, respectively.

The fermionic pseudomode Fock state is defined as $|{\bf n}\rangle = | \cdots, n_{\alpha_k p_k \sigma_k},\cdots \rangle$, where $n_{\alpha_k p_k \sigma_k} \in \{0,1\}$ specifies the occupation of the pseudofermion associated with the $p_k$-th exponential term in \Eq{FermiBCF_exp}, corresponding to the sign $\sigma_k$, and electrode $\alpha_k$. When $n_{\alpha_k p_k\sigma_k}=1$, the pseudofermion is occupied. Otherwise, when $n_{\alpha_k p_k\sigma_k}=0$, this effective electronic level is empty. Similarly, the bosonic pseudomode Fock state is written as $|{\bf m}\rangle = |\cdots, m_{ \theta_l p_l},\cdots \rangle$, where $m_{\theta_l p_l}$ is a non-negative integer, denoting the occupation number of the pseudoboson related to the $p_l$-th term in \Eq{BosonBCF_exp}, for the bath coupled to the molecular vibration or a cavity mode. For notational simplicity, we hereafter denote $n_{\alpha_k p_k \sigma_k}\rightarrow n_k$ and $m_{\theta_l p_l}\rightarrow m_l$.

The system is represented in twin space, in which the basis is doubled and purified from a bra and a ket in Hilbert space:
\begin{equation}
    |{\bf s}, {\bf s}'\rangle = |s^{}_{\rm e}, s'_{\rm e}, s^{}_{\rm v}, s'_{\rm v}, s^{}_{1_{\mathrm{c}}}, s'_{1_{\mathrm{c}}},s^{}_{2_{\mathrm{c}}}, s'_{2_{\mathrm{c}}}, \cdots\rangle.
\end{equation}
Here, $s^{}_{\rm e}$ and $s'_{\rm e}$ specify the occupation of the molecular electronic orbital; $s^{}_{\rm v}$ and $s'_{\rm v}$ denote the molecular vibrational states; $s^{}_{i_{\rm c}}$ and $s'_{i_{\mathrm{c}}}$ represent the states of the $i_{\mathrm{c}}$-th cavity mode. 

The super-Hamiltonian $\mathcal{H}$ in \Eq{SchroedingerEquation} is non-Hermitian, and its explicit form depends subtly on the bath characteristics, the pole decomposition scheme in \Eq{BCF_exp}, as well as the specific definition of the auxiliary density operators. In this work, we employ Lorentzian spectral density functions, 
\begin{subequations}
\label{Lorentzian}
\begin{equation}
     \Gamma_{\alpha}(\epsilon) = \frac{(\lambda_{\alpha}^{\rm F}\Omega_{\alpha})^2}{(\epsilon-\mu_{\alpha})^2+\Omega_{\alpha}^2},
\end{equation}
\begin{equation}
    J_{\theta}(\omega) =  \frac{2(\lambda_{\theta}^{\rm B})^2\omega\Omega_{\theta}}{\omega^2+\Omega_{\theta}^2},
\end{equation}
\end{subequations}
together with the Pad\'e pole decomposition scheme.\cite{Hu_2010_JCP_p101106}  Under these conditions, the resulting super-Hamiltonian takes the form 
\begin{equation}
\label{superHamiltonian}
\begin{split}
    \mathcal{H} = &\hat{H}_s-\tilde{H}_{s} +\sum_{\theta} (\lambda^{\rm B}_{\theta})^2 (h_{\theta}^2(\hat{x}_{\theta}) - h_{\theta}^2(\tilde{x}_{\theta})) + \mathcal{L}_{\rm CAP}\\
    &-i\sum_{k} \gamma^{\rm F}_{\alpha_k p_k\sigma_k} a^{<,+}_{k}a^{<,-}_k  - \sum_k \lambda^{\rm F}_{\alpha_k} \left(\hat{d}^{\bar{\sigma}_k}a^{<,-}_k-\tilde{d}^{\bar{\sigma}_k}a^{>,-}_k\right) \\
    &- \sum_k \lambda^{\rm F}_{\alpha_k} \left(\eta^{\rm F}_{\alpha_k p_k \sigma_k}\hat{d}^{\sigma_k}a^{<,+}_k-\eta^{\rm F *}_{\alpha_k p_k\bar{\sigma}_k}\tilde{d}^{\sigma_k}a^{>,+}_k\right) \\
    & -i\sum_l \gamma^{\rm B}_{\theta_lp_l} b_l^{+}b_l^{-} -\sum_l \lambda^{\rm B}_{\theta_{l}} (h_{\theta_l}(\hat{x}_{\theta_{l}})-h_{\theta_l}(\tilde{x}_{\theta_{l}}))b_l^{-} \\
    & - \sum_l \lambda^{\rm B}_{\theta_{l}} (\eta^{\rm B}_{\theta_{l}p_l}h_{\theta_l}(\hat{x}_{\theta_{l}})-\eta^{\rm B *}_{\theta_{l}p_l}h_{\theta_l}(\tilde{x}_{\theta_{l}}))b_l^{+}. 
\end{split}
\end{equation}
Here, $\hat{H}_S$ and $\tilde{H}_S$ are the system super-Hamiltonian operators in twin space, dilated from the physical Hamiltonian $H_S$ in \Eq{SystemHamiltonian}. Specifically, the superoperators $\hat{d}^{\pm}$ and $\tilde{d}^{\pm }$ originate from $d^{\pm}$, i.e., $\hat{d}^{\pm}=d^{\pm}\otimes I_e$ and  $\tilde{d}^{\pm}=I_e\otimes d^{\mp}$. They act on $|s_{\rm e}\rangle$ and $|s'_{\rm e}\rangle$, respectively. Using the Jordan-Wigner transformation,\cite{Jordan_1928_ZP_p631} these operators can be represented in terms of spin operators. Analogous definitions apply to the vibrational and cavity-mode coordinates: $\hat{x}_{\rm m}$ and $\tilde{x}_{\rm m}$ are related to $x_{\rm m}$ and act on $|s_{\rm v}\rangle$ and $|s'_{\rm v}\rangle$, respectively; $\hat{x}_{i_{\rm c}}$ and $\tilde{x}_{i_{\rm c}}$ originate from $x_{i_{\rm c}}$ and they act on $|s_{i_{\rm c}}\rangle$ and $|s'_{i_{\rm c}}\rangle$, respectively. The term $\mathcal{L}_{\rm CAP}$ corresponds to a complex abosrbing potential used to describe the bond rupture effect, which will be explained in detail later.

The operators $a^{\lessgtr,  \pm}_k$ and $b^{\pm}_l$ in \Eq{superHamiltonian} act on the pseudomode Fock state $|\bf n\rangle$ and $|\bf m\rangle$, respectively. Specifically, we define
\begin{subequations}
\begin{equation}
    a_k^{\lessgtr,+} |{\bf n}\rangle = (-1)^{\sum_{j\lessgtr k}n_j}\sqrt{1-n_k} |{\bf n}+{\bf 1}_k\rangle,
\end{equation}   
\begin{equation}
    a_k^{\lessgtr,-} |{\bf n}\rangle = (-1)^{\sum_{j\lessgtr k}n_j}\sqrt{n_k} |{\bf n}-{\bf 1}_k\rangle,
\end{equation} 
\begin{equation}
    a_k^{\lessgtr,+} a_k^{\lessgtr,-} |{\bf n}\rangle = n_k |{\bf n}\rangle.
\end{equation} 
\end{subequations}
where $|{\bf n}\pm {\bf 1}_k\rangle= |\cdots, 1-n_k, \cdots \rangle$. These pseudofermion operators commute with all system operators. For the bosonic pseudomodes, the ladder operators satisfy
\begin{subequations}
    \begin{equation}
        b_l^+ |{\bf m}\rangle = \sqrt{m_l+1}|{\bf m}+{\bf 1}_l\rangle,
    \end{equation}
    \begin{equation}
        b_l^- |{\bf m}\rangle = \sqrt{m_l}|{\bf m}-{\bf 1}_l\rangle,
    \end{equation}
    \begin{equation}
       b_l^+ b_l^- |{\bf m}\rangle = m_l|{\bf m}\rangle,
    \end{equation}
\end{subequations}
with $|{\bf m} \pm {\bf 1}_l\rangle = |\cdots, m_l\pm 1, \cdots\rangle$.

To enable numerically exact and efficient dynamic simulations, the high-dimensional coefficient tensor $C_{\bf{s}, \bf{s}'}^{ \bf{n}, \bf{m}}(t)$ appearing in the extended wave function \Eq{ExtendedWaveFunction} is decomposed into a tree tensor network state (TTNS),\cite{Shi_Phys.Rev.A_2006_p22320}\cite{Ke_2023_JCP_p211102}
\begin{equation}
    C_{\bf{s}, \bf{s}'}^{ \bf{n}, \bf{m}}(t) = {\rm Contr}\left\{ T^{[1]}_{s^{}_{\rm e}}T^{[2]}_{s'_{\rm e}}  \cdots T^{[k]}_X\cdots \right\},
\end{equation}
which is a network of interconnected low-rank tensors $T_{X}$, and ${\rm Contr} \{\}$ denotes the contraction over all virtual indices. An illustrative example for the model system with seven discrete cavity modes is shown in \Fig{fig1:schematic}~c). The corresponding super-Hamiltonian $\mathcal{H}$ can be likewise decomposed into a tree tensor network operator (TTNO) that shares the same network topology as the TTNS.\cite{Li_J.Chem.Phys._2024_p54116,Cakir_2025_PRB_p35101} Time evolution is carried out based on a time-dependent variational principle (TDVP)-based propagation scheme,\cite{Bauernfeind_SciPostPhysics_2020_p24,Ceruti_SIAMJ.Numer.Anal._2021_p289} following the algorithm described in Ref.\onlinecite{Ke_2023_JCP_p211102}. Owing to the flexibility and compactness of the TTNS ansatz, this framework allows us to efficiently converge the fully correlated, large-scale molecule-cavity-enviroment wavefunction. 

\subsection{Simulation Details}
Using the method described above, we are able to treat all DoFs--photons, molecular vibrations, electrons, and their associated environmental modes—on an equal, fully quantum-mechanical footing. All system observables and system–bath transport properties can be obtained directly from the extended wavefunction $|\Psi(t)\rangle$.

For the simulations presented in this work, we adopt Morse potential surface parameters $D_{\rm g}=D_{\rm e}=1.4\,{\mathrm eV}$, $a_{\rm g}=a_{\rm e}=3\text{\r{A}}$ for both the neutral and charged states. The equilibrium position of the neutral state is set to $x_{\rm m}^{\rm g, 0}=1.8\,\text{\r A}$, while the charged PES is displaced to  $x_{\rm m}^{\rm e, 0}=1.9\,\text{\r A}$. The charging energy is taken to be $E=1\,\mathrm{eV}$. A molecular mass of $M_{\rm S}=1\,\mathrm{amu}$ is used throughout.  

The reactive vibrational mode is described in the potential-optimized discrete-variable representation (DVR).\cite{Colbert_1992_JCP_p1982,Echave_1992_CPL_p225}  Specifically, we employ the sine-DVR to represent the operator $\frac{p_{\rm m}^2}{2M_{\rm S}}+V_{g}(x_{\rm m})$ on the grids of $x_{\rm m}$ spanning from $x_{\rm m}^{\rm min}=1.4\,\text{\r A}$ to $x_{\rm m}^{\rm max}=5\,\text{\r A}$. Then, we diagonalize this representation, sort in ascending order and truncate the matrix to yield the lowest $N_{\rm v}$ vibrational eigenenergies in the neutral electronic state and the associated $N_{\rm v}$ eigenvectors $U=\{{\bf v}_1, \cdots, {\bf v}_{N_{\rm v}}\}$. For the chosen parameters, eight discrete vibrational levels lie below the dissociation threshold, as shown in \Fig{fig1:schematic}~b), with inequivalent level spacings that decrease by approximately $300\,\mathrm{cm}^{-1}$ with increasing vibrational quantum number. States above the dissociation limit are quasi-continuous and unbounded. However, under a finite bias voltage, highly excited vibrational states beyond a certain level contribute negligibly to the reaction dynamics, as they are hardly populated. We therefore systematically increase the truncated basis size $N_{\rm v}$  until convergence. All functions of the reaction coordinate $f(x_{\rm m})$--including the configuration-dependent modification to the charging energy $V_e(x_{\rm m})-V_g(x_{\rm g})$, the molecule-bath coupling operator $h_{\rm m}(x_{\rm m})$, and the molecular dipole function $u(x_{\rm m})$--are also represented in this vibrational eigenbasis.  Photonic modes are harmonic and are represented in their eigenbasis, truncated to the lowest $N_{\rm c}$ levels. For all parameter sets considered, we use $N_{\rm m}=30$ and $N_{\rm c}=6$.

To suppress finite-size artifacts and capture bond rupture, we introduce a complex absorbing potential (CAP)\cite{Erpenbeck_J.Chem.Phys._2019_p191101}
\begin{equation}
    V_{\rm CAP}(x_{\rm m}) = -i \xi (x_{\rm m}-x_{\rm m}^{\ddagger})^4\Theta(x_{\rm m}-x_{\rm m}^{\ddagger})
\end{equation}
with $\xi=5\,\mathrm{eV}/{\text{\r A}^4}$ and $x_{\rm m}^{\ddagger}=3.4\,\text{\r A}$.

The CAP absorbs outgoing wavepackets beyond 
$x_{\rm m}^{\dagger}$, as indicated by the Heaviside function $\Theta(x_{\rm m}-x_{\rm m}^{\ddagger})$. The corresponding term in the super-Hamiltonian of \Eq{superHamiltonian} is given explicitly by
\begin{equation}
    \mathcal{L}_{\rm CAP} = V_{\rm CAP}(\hat{x}_{\rm m})+V_{\rm CAP}(\tilde{x}_{\rm m}). 
\end{equation}
The CAP induces a loss of population from the reactive region, which we identify with the dissociation probability. The cumulative loss is computed as 
\begin{equation}
Q_{\rm loss}(t) = \int_0^t  Q_{\rm CAP}(\tau) \mathrm{d}\tau
\end{equation}
with the instantaneous loss rate
\begin{equation}
   Q_{\rm CAP}(\tau) = \int_{x_{\rm m}^{\rm min}}^{x_{\rm m}^{\rm max}} \langle \mathcal{I}_0|\sqrt{2i V_{\rm CAP}(\hat{x}_{\rm m})}\otimes\sqrt{2i V_{\rm CAP}(\tilde{x}_{\rm m})}|\Psi(\tau)\rangle \mathrm{d}x_{\rm m}. 
\end{equation}
Here, $|\mathcal{I}_0\rangle = \sum_{\mathbf s} | \mathbf{s}, \mathbf{s}\rangle\otimes |\mathbf{n=0}\rangle \otimes |\mathbf{m=0}\rangle$ is an extended wavefunction that, when contracted with $|\Psi(t)\rangle$, performs the trace over the system DoFs while projecting all environmental pseudomodes onto their vacuum states. This definition of the dissociation probability $Q_{\rm loss}(t)$ is formally equivalent to introducing a distant dissociation site $x_{\infty}$ far from the equilibrium together with an associated Lindblad-type escape channel, as employed in our previous works.\cite{Erpenbeck_Phys.Rev.B_2020_p195421,Ke_J.Chem.Phys._2021_p234702,Ke_2023_JCP_p24703}

Bond rupture characterized in this manner is irreversible and the reaction therefore proceeds unidirectionally. At long times, we observe in all simulations that the surviving population $1-Q_{\rm loss}(t)$ decays exponentially, enabling the extraction of a constant reaction rate,
\begin{equation}
    k = \lim_{t\rightarrow t_{\rm plateau}}k(t) =  \lim_{t\rightarrow t_{\rm plateau}} \frac{d\ln(1-Q_{\rm loss}(t))}{dt}. 
\end{equation}
Reprensentative examples demonstrating the convergence of $k(t)$ to a well-defined plateau, from which the reaction rate is determined, are provided in the supplementary material.  We denote the cavity-modified rate by $k_{\rm c}$, and the rate in the absence of cavity photons as $k_{\rm o}$.

Any other system observables, such as the electronic-state populations $P_g(t)$ and $P_e(t)$, or the populations of different vibrational levels $P_{v_i}(t)$, can be obtained from the extended wave function via  $\langle \mathcal{I}_0|\hat{O}|\Psi(t)\rangle$, where $O$ the corresponding operator acting in the system subspace. 

Unless otherwise stated, we focus on symmetric molecule-electrode coupling conditions, with $\lambda_{L}^{\rm F}=\lambda_{R}^{\rm F}=0.5\,\mathrm{eV}$ under a positive bias voltage $\Phi$. In this case, the left electrode serves as the source and the right electrode as the drain. Extensions to asymmetric coupling scenarios, along with detailed discussions, are presented in the supplementary material. The electronic current running from electrode $\alpha$ into the molecule is given by
\begin{equation}
    I_{\alpha}(t) = -i\sum_k \sigma_k\lambda_{\alpha}^{\rm F}\delta_{\alpha_k, \alpha} \langle \mathcal{I}_{\mathbf{n}=\mathbf{ 1}_k}|\hat{d}^{\sigma_k}|\Psi(t)\rangle, 
\end{equation}
where $\delta_{\alpha_k, \alpha}$ is the Kronecker delta and $|\mathcal{I}_{\mathbf{n}={\mathbf{1}_k}}\rangle = \sum_{\mathbf s} | \mathbf{s}, \mathbf{s}\rangle\otimes |\mathbf{n}=\mathbf{1}_k\rangle \otimes |\mathbf{m=0}\rangle$ is the tensor product of the identity wavefunction in the system subspace, the single excitation state in the $k$-th pseudofermion, and the ground state for all other pseudomodes. The net current is then obtained by $I_e(t) = I_{L}(t)-I_{R}(t)$. 

Both fermionic and bosonic environments are set at room temperature, $T=300\,\mathrm{K}$. The characteristic frequencies in the Lorentzian spectral density functions of \Eq{Lorentzian} are set to $\Omega_{L}=\Omega_R=1\,\mathrm{eV}$ for the electrodes, $\Omega_{\rm m}=200\, \mathrm{cm}^{-1}$ for the molecular bath, and $\Omega_{i_{\rm c}}=1000\,\mathrm{cm}^{-1}$ for all cavity baths. With these parameters, the exponential decomposition of the bath correlation functions in \Eq{BCF_exp} converges with $P^{\rm F}=10$ (11 pseudofermions) and $P^{\rm B}=3$ (4 pseudobosons). 

Time evolution begins with the molecular wavepacket in the neutral electronic state. The photonic and vibrational populations are initialized according to their respective Boltzamn distributions, $e^{-\beta H_{\rm c}}/\mathrm{tr}\{e^{-\beta H_{\rm c}}\}$ ($H_{\rm c}$ is the free cavity Hamiltonian) and $e^{-\beta H_g}/\mathrm{tr}\{e^{-\beta H_{g}}\}$ ($H_{\rm g}$ is the molecular vibrational Hamiltonian in the neutral electronic state). Each bath is initially uncorrelated with the system and prepared in its own thermal equilibrium state, which in the HEOM framework corresponds to initializing all pseudomodes in their vacuum states. The time step for propagation in all simulations is $\Delta t=0.5\,\text{fs}$, and the maximum bond dimension for the TTNS decomposition ansatz is $D_{\rm max}=40$. 

\section{\label{sec:results}Results}
We begin by analyzing the non-equilibrium bond dissociation dynamics in the absence of optical confinement, where the molecule is coupled to two macroscopic electrodes and embedded in a dissipative solvent environment. This configuration corresponds to the standard setting of molecular junctions and scanning tunneling microscopy (STM) experiments, in which electronic transport drives molecular heating and may even induce bond rupture. Establishing this baseline is essential, as it clarifies the transport-induced dissociation mechanisms and provides the reference rate against which cavity-induced modifications will be assessed.

With this foundation, we then introduce the confined electromagnetic field and systematically examine how molecule-cavity interactions affect the reaction dynamics. We first focus on a single cavity mode, allowing us to track how the dissociation rate varies as a function of the cavity frequency and to identify resonant rate suppressions that arise from photon-assisted vibrational relaxations. Building on these insights, we next explore the possibility of cavity-induced vibrational cooling, wherein the multi-mode cavity facilitates an efficient vibrational ladder-descending process that counteracts current-induced heating. Such cooling mechanisms offer a potential route to stabilize reactive intermediates and suppress bond breaking in a current-carrying molecular junction under strong electronic nonequilibrium. 

\begin{figure}
\centering
 \begin{minipage}[c]{0.45\textwidth}
    \raggedright a) 
    \includegraphics[width=\textwidth]{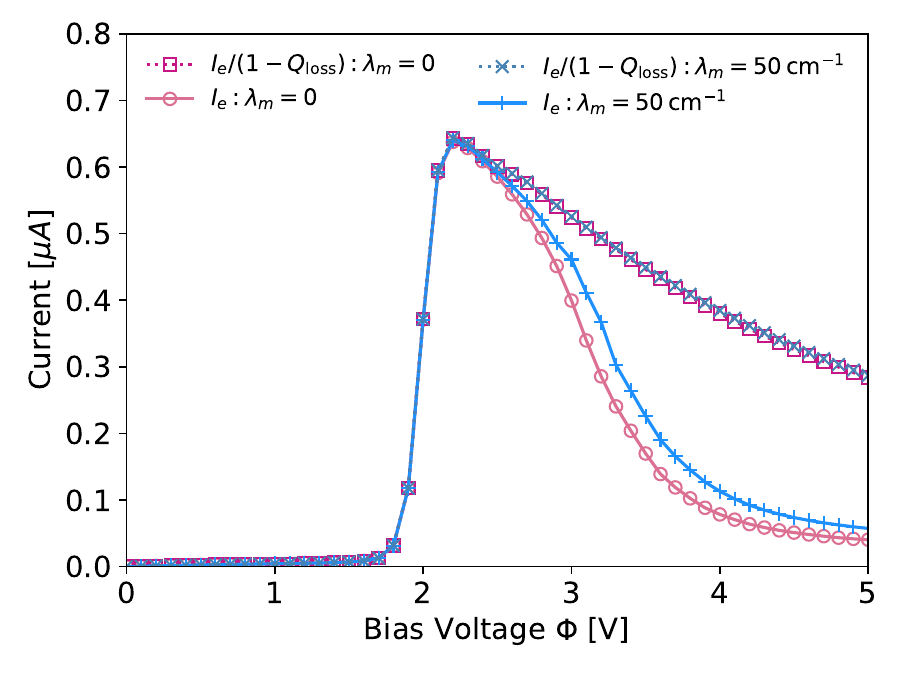}
 \end{minipage}
  \begin{minipage}[c]{0.45\textwidth}
    \raggedright b) 
    \includegraphics[width=\textwidth]{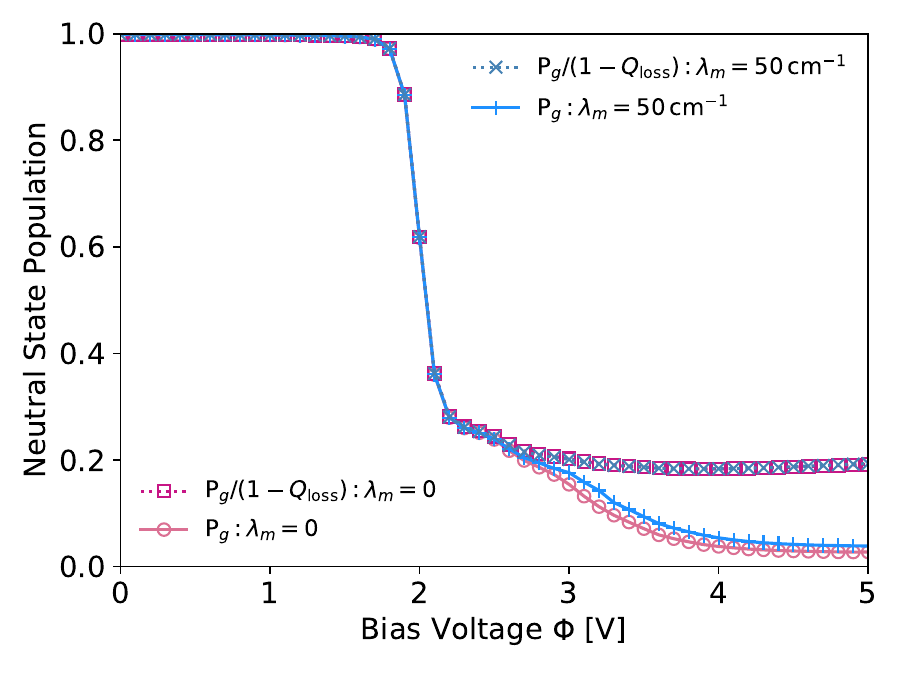}
 \end{minipage}
  \begin{minipage}[c]{0.45\textwidth}
    \raggedright c) 
    \includegraphics[width=\textwidth]{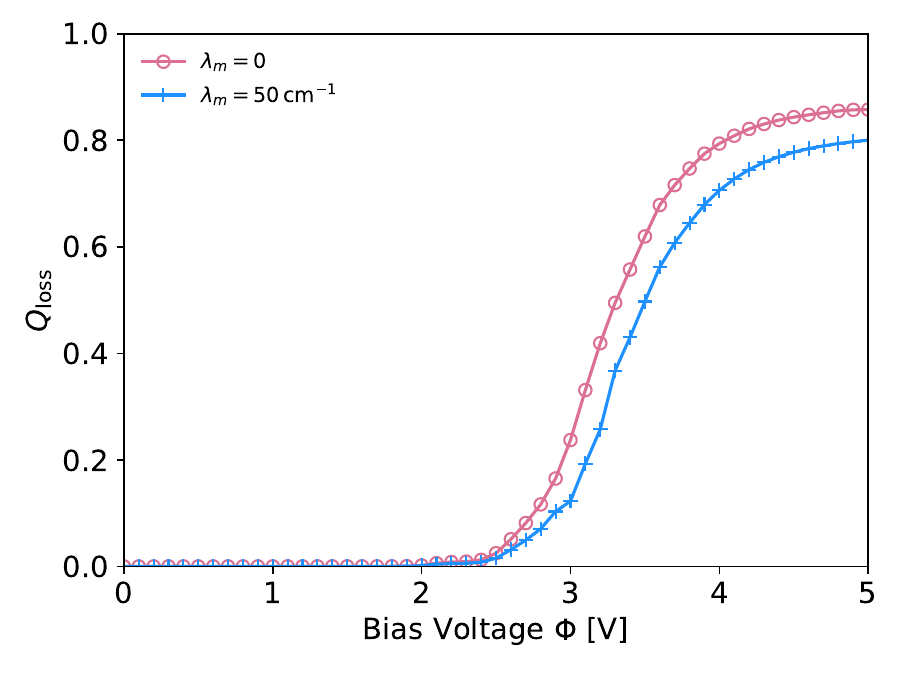}
 \end{minipage}
\caption{Electronic current $I_e$ and its value rescaled value by the surviving population $1-Q_{\rm loss}$ in panel a); Population of the neutral electronic state $P_g$ and its rescaling $P_g/(1-Q_{\rm loss})$ in panel b); Dissociation probability $Q_{\rm loss}$ in panel c).
All quantities are shown as functions of the applied bias voltage $\Phi$, varied from 0 to 5~V. These observables are evaluated  at time $t=10$~ps. The pink and blue lines correspond, respectively, to the cases without and with coupling to a dissipative bosonic bath ($\lambda_{\rm m}=0$ and $\lambda_{\rm m}=50\,\mathrm{cm}^{-1}$), representing solvent or surface phonon-induced vibrational relxation. } \label{fig2:outsidecavity}
\end{figure}

\begin{figure}
\centering
 \begin{minipage}[c]{0.45\textwidth}
    \includegraphics[width=\textwidth]{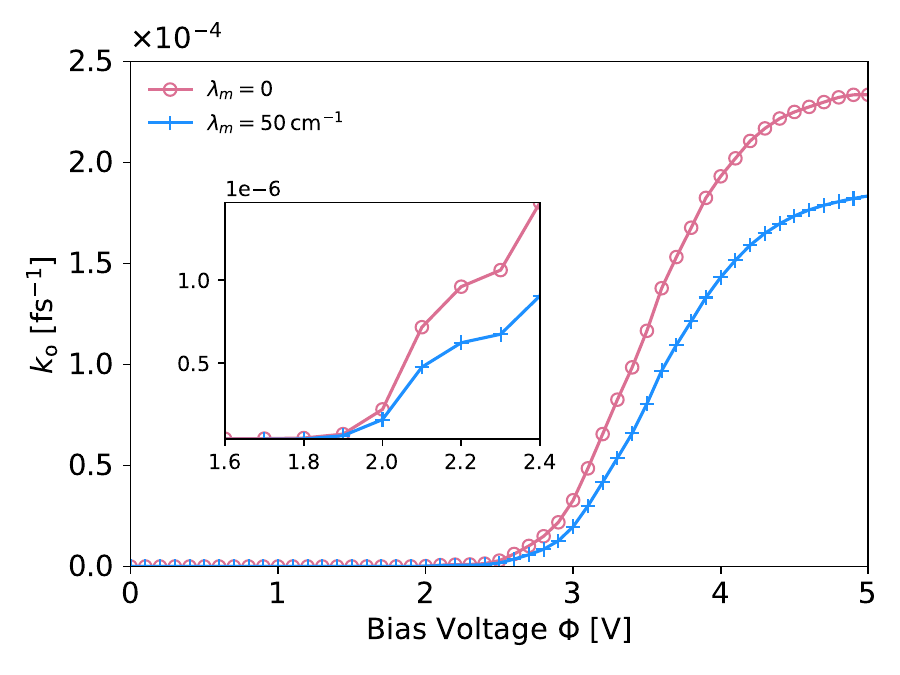}
 \end{minipage}
\caption{Dissociation rate $k_{\rm o}$ in the radiation-free case as a function of the applied bias voltage $\Phi$. The inset provides a close-up view of the crossover region between non-resonant and resonant electronic transport. The pink curve corresponds to $\lambda_{\rm}=0$, where the dissipation is caused solely by electron-hole pair generation in the electrodes. The blue curve corresponds to $\lambda_{\rm m}=50\,\mathrm{cm}^{-1}$, which introduces an additional relaxation pathway through coupling to a dissipative bosonic bath.} \label{fig3:outsidecavityrates}
\end{figure}

\subsection{\label{subsec:modelI}Outside Optical Cavity}
Under an applied bias voltage, electrons are continuously injected into and transferred out of the molecule through the source and drain electrodes. Each charging and discharging event is accompanied by the absorption or emission of vibrational quanta, with the strength of these vibronic transitions in our model determined primarily by the displacement between the minima of the neutral and the shifted charged PESs, $x_{\rm m}^{e,0}-x_{\rm m}^{g,0}$.  Consequently, electronic transport drives the coupled vibrational DoF out of equilibrium. Over time, part of the nuclear wavepacket is promoted toward the barrier region and eventually crosses into the dissociative continuum, resulting in irreversible bond rupture. 

The dominant dissociation mechanism depends sensitively on three factors: the applied bias voltage, the molecule–electrode coupling, and the vibronic coupling strength. When the vibronic coupling is weak, dissociation proceeds primarily through stepwise vibrational ladder climbing, which requires a fully quantum-mechanical description of the nuclear motion. In contrast, under strong vibronic coupling combined with a large bias voltage, direct transitions from a deep bound vibrational state into the unbound continuum become both energetically accessible and kinetically favorable, providing a faster dissociation pathway. These mechanisms and their corresponding parameter regimes have been analyzed in detail in our previous work.\cite{Ke_J.Chem.Phys._2021_p234702} Broadly speaking, dissociation accelerates dramatically at high bias voltages, which is the main cause of junction failure in molecular electronics. 

\Fig{fig2:outsidecavity} displays the electronic current, the population of the neutral electronic state, and the dissociation probability at $t=10\,\mathrm{ps}$ as functions of the bias voltage $\Phi$ over the range 0-5~V for two solvent coupling strengths: $\lambda_{\rm m}=0$ and $50\,\mathrm{cm}^{-1}$. Full time-dependent traces of these observables are provided in the supplementary material. 

For the parameters considered in this work, the molecule is highly stable under thermal equilibrium conditions (i.e., $\Phi=0$ and detached from the electrodes), remaining intact far exceeding our simulation time window, because the dissociation barrier is much larger than the thermal energy $k_{B}T$. However, as the bias voltage increases and once the chemical potential of the left electrode approaches the molecular charging energy $E=1\,\mathrm{eV}$, electron transport undergoes a crossover from non-resonant cotunneling to resonant sequential tunneling. This transition is manifested by (i) a sharp rise in the electronic current and (ii) a rapid drop of the neutral-state population near $\Phi=2\,\mathrm{V}$, as shown in \Fig{fig2:outsidecavity}~a) and b) respectively. 

The onset of significant dissociation occurs at slightly larger voltages (see \Fig{fig2:outsidecavity}~c)), when sufficient vibrational energy is accumulated through current-induced heating and high-lying vibrationally excited states become appreciably populated to overcome the barrier. Once dissociation sets in, the fraction of intact molecules decreases steadily. Because we assume that dissociated molecules become non-conductive, the observed current drops at high biases. Thus, the current exhibits a falloff at voltages above $2.2\,\mathrm{V}$, due to the increasing probability of bond rupture. The current associated with the surviving molecules can be recovered by rescaling the measured current as $I_e(t)/(1-Q_{\rm loss}(t))$. Interestingly, even after this rescaling, the intrinsic current–voltage characteristics still exhibit prominent negative differential resistance. However, the rescaled electronic state populations, such as $P_{g}/(1-Q_{\rm diss})$, reach a plateau at high voltages, indicating that molecules that have not dissociated settle into a steady-state electronic distribution that is nearly insensitive to further increases in the bias voltage.

The dissociation rates in electromagnetic free space, $k_{\rm o}$, obtained from the long-time exponential decay of the surviving population, are shown in \Fig{fig3:outsidecavityrates}.  The voltage dependence of $k_{\rm o}$ closely mirrors the broadened, step-like profile of the dissociation probabilities. When $\lambda_{\rm m}=0$, vibrational cooling arises exclusively from electron-hole pair generations in the electrodes,\cite{Haertle_2013_PSSB_p2365} which extract excessive vibrational energy from the molecule during charge transport.  Introducing coupling to a solvent or electrode phonons ($\lambda_{\rm m}\neq 0$) adds additional relaxation channels that accelerate vibrational energy dissipation. Nevertheless, even with appreciable solvent-induced damping, dissociation remains substantial in the resonant transport regime, indicating that generic solvent-induced vibrational relaxation alone is insufficient to counteract transport-driven heating.

Furthermore, we emphasize that a key limitation of the solvent-induced relaxation is its non-selective character: it damps vibrational excitations broadly with little frequency specificity. Such broadband dissipation does not preferentially cool the vibrational modes that are most strongly driven out of equilibrium by electronic transport. By contrast, a confined electromagnetic field--such as a cavity photon mode or a localized plasmonic resonance--can mediate mode-selective and frequency-resolved energy transfer between targeted molecular vibrations and the photonic environment. This raises the possibility of stabilizing molecular junctions not by indiscriminately increasing overall dissipation, but by selectively and effectively enhancing cooling pathways that directly counteract current-induced vibrational excitation in the reactive bond responsible for junction failure. In the following sections, we investigate whether such cavity-assisted cooling can occur and to what extent the confined light field can suppress dissociation and thereby enhance molecular stability under bias.

\begin{figure}
\centering
 \begin{minipage}[c]{0.45\textwidth}
     \raggedright a) 
    \includegraphics[width=\textwidth]{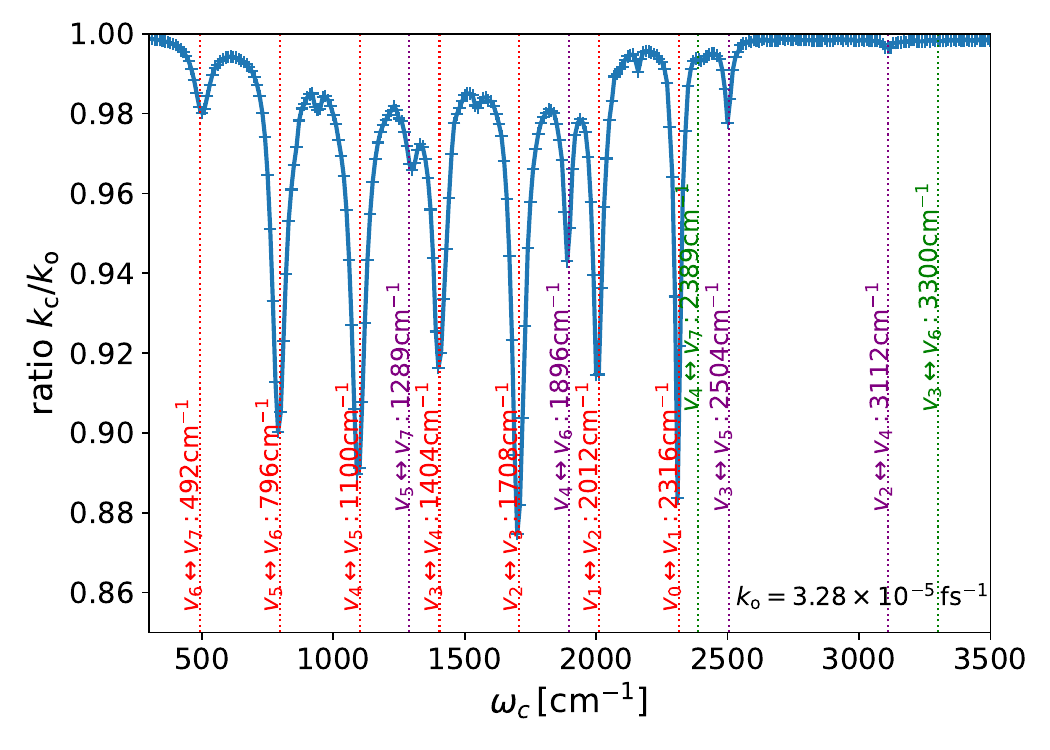}
 \end{minipage}
  \begin{minipage}[c]{0.45\textwidth}
      \raggedright b) 
    \includegraphics[width=\textwidth]{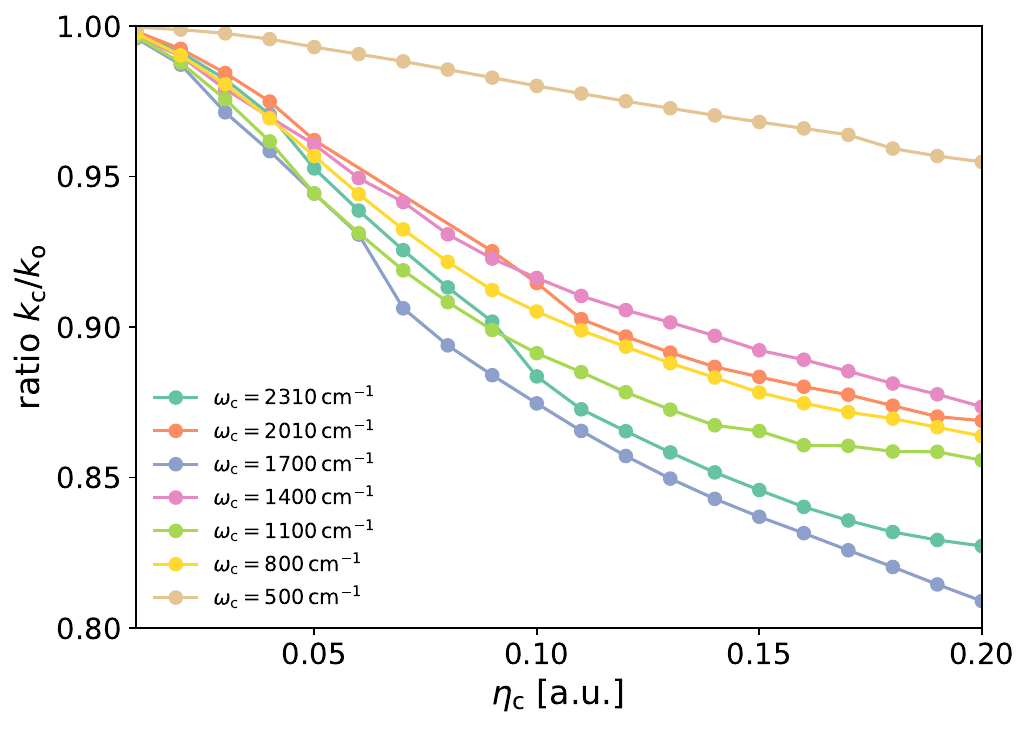}
 \end{minipage}
\caption{a) Ratio of dissociation rates inside and outside a single-mode cavity, $k_{\rm c}/k_{\rm o}$, as a function of the cavity frequency $\omega_{\rm c}$ spanning from $300$ to $3500\,\mathrm{cm}^{-1}$ in steps of $10\,\mathrm{cm}^{-1}$. The light-matter coupling strength is fixed at $\eta_{\rm c}=0.1$~a.u. The red, purple, and green dotted vertical lines mark the positions of distinct one-quantum ($v_i\leftrightarrow v_{i+1}$, two-quantum ($v_i\leftrightarrow v_{i+2}$, and three-quantum ($v_i\leftrightarrow v_{i+3}$) vibrational transitions, respectively. b) Cavity-induced rate modification ratio $k_{\rm c}/k_{\rm o}$ plotted as a function of $\eta_{\rm c}$ for seven different cavity frequencies chosen to be in close resonance with discrete vibrational transition energies along the vibrational energy ladder up to the dissciation barrier.  The electronic dipole function is given by $u(x_{\rm m})=x_{\rm m}e^{-x^2_{\rm m}/a_{u}^2}$ with $a_{u}=2.5\,\text{\r{A}}$. Dissipation into the bosonic bath is neglected, i.e., $\lambda_{\rm m}=0$. The bias voltage is fixed at $\Phi=3$~V.  } \label{fig4:insidesidecavityratio1}
\end{figure}

\begin{figure}
\centering
 \begin{minipage}[c]{0.45\textwidth}
     \raggedright a) 
    \includegraphics[width=\textwidth]{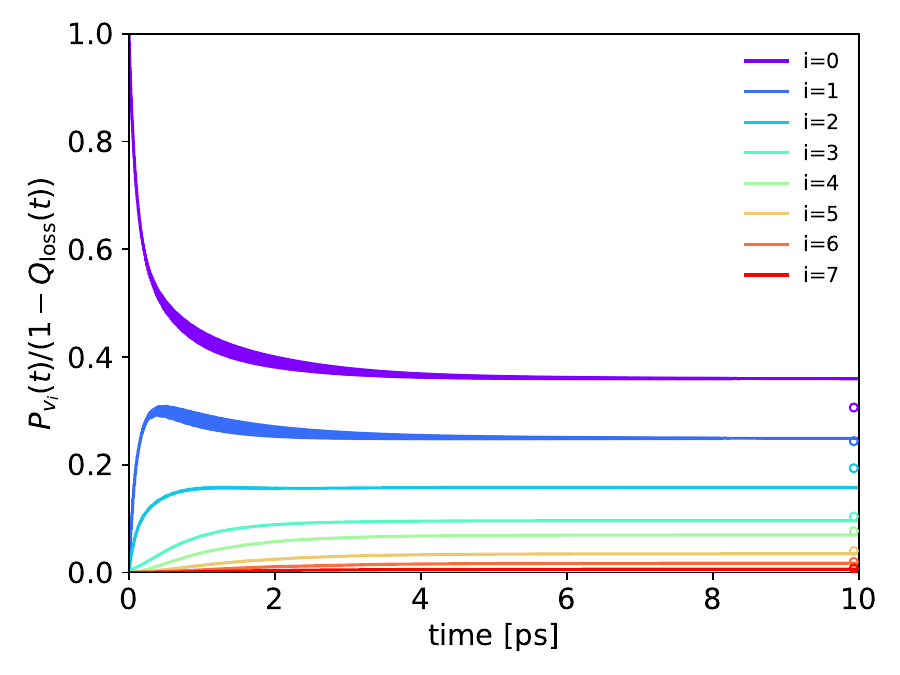}
 \end{minipage}
  \begin{minipage}[c]{0.45\textwidth}
      \raggedright b) 
    \includegraphics[width=\textwidth]{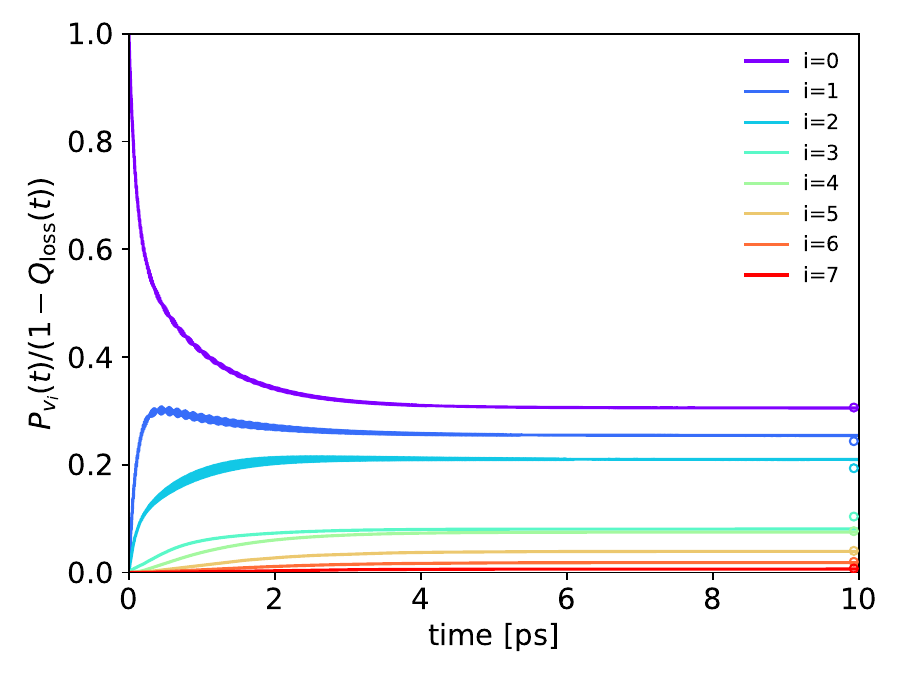}
 \end{minipage}
\caption{Time evolution of the vibrational population distribution $P_{v_i}(t)$ for bound vibrational states below the dissociation barrier at the bias voltage of $\Phi=3$~V, shown for two different cavity frequencies: $\omega_{\rm c}=2310\,\mathrm{cm}^{-1}$ in panel a)  and $\omega_{\rm c}=2010\,\mathrm{cm}^{-1}$ in panel b).  All populations are rescaled by the surviving probability $1-Q_{\rm loss}(t)$. The dipole function is $u(x_{\rm m})=x_{\rm m}e^{-x^2_{\rm m}/a_{u}^2}$ with $a_{u}=2.5\,\text{\r{A}}$. Dissipation into the bosonic bath is neglected $(\lambda_{\rm m}=0)$. For comparison, the corresponding long-time steady vibrational populations outside the cavity are plotted as circles along the right axis. 
} \label{fig5:vibpop}
\end{figure}

\begin{figure}
\centering
  \begin{minipage}[c]{0.5\textwidth}
    \includegraphics[width=\textwidth]{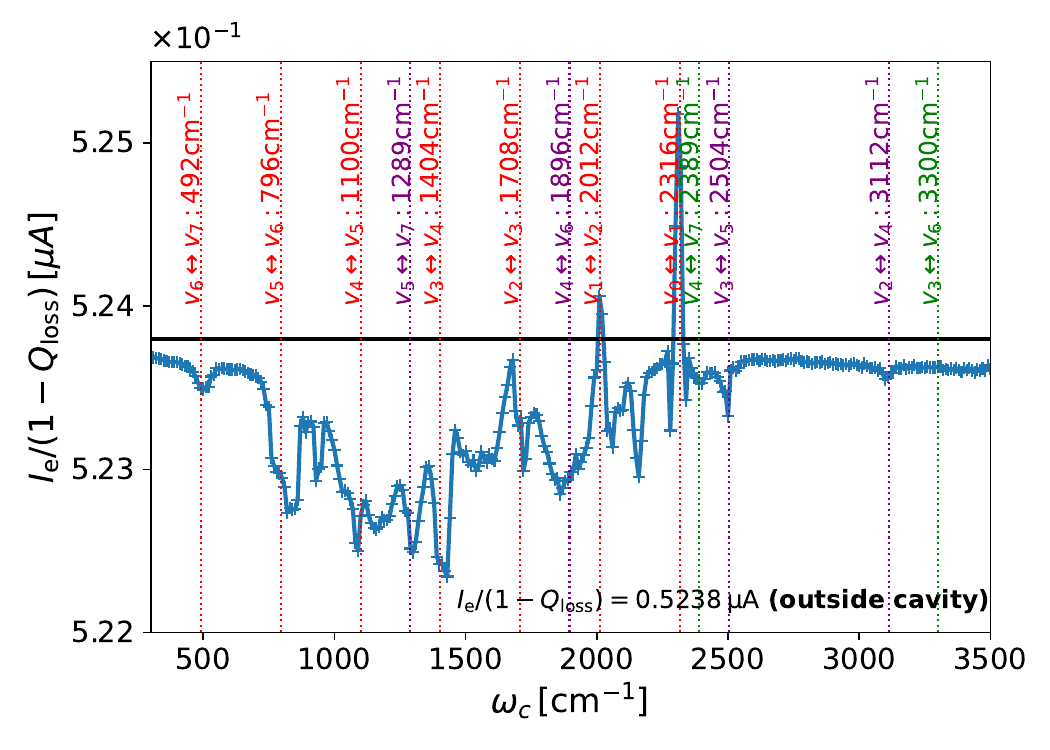}
 \end{minipage}
\caption{Rescaled electronic current, $I_e/(1-Q_{\rm loss})$, obtained at $t=10$~ps inside a single-mode cavity as a function of the cavity frequency $\omega_c$. The ccorresponding urrent outside the cavity is indicated by the thick horizontal black line. The dipole function is $u(x_{\rm m})=x_{\rm m}e^{-x^2_{\rm m}/a_{u}^2}$ with $a_{u}=2.5\,\text{\r{A}}$. Other parameters are $\eta_{\rm c}=0.1$~a.u., $\Phi=3$~V, and $\lambda_{\rm m}=0$. } \label{fig6:insidesidecavitycurrent}
\end{figure}

\subsection{\label{subsec:modelI}Inside the confined light field}
When a molecule is placed inside a highly confined electromagnetic environment, such as an optical cavity or plasmonic nanogap, it interacts with the enhanced local light field. In the strong coupling regime, this interaction hybridizes molecular and photonic DoFs, forming new dressed eigenstates--polaritons--that can reshape the molecular energy landscape and redirect dynamical pathways. Such hybridization underpins a variety of intriguing phenomena, such as altered reaction rates, the opening of long-range energy transfer channels, and changes in nonadiabatic transport processes. 

The magnitude of the light-matter interaction depends on several factors, including the molecular transition dipole moment, the photon frequency, the mode volume of the confined field, and the permittivity of the surrounding medium. In principle, the relevant dipole moment as a function of nuclear geometry can be extracted from ab initio electronic structure calculations. However, for the present study--which focuses on an anharmonic reactive vibration and aims to explore generic cavity effects independent of molecular specificity, we adopt the following parametrized dipole function:\cite{Hernandez_2019_JCP_p144116,Triana_2020_JCP_p234111,Fischer_2021_JCP_p104311}
\begin{equation}
    u(x_{\rm m}) = (x_{\rm m } - c_u)e^{-x_{\rm m}^2/\sigma_u^2},
\end{equation}
where the parameters $c_u$ and $\sigma_u$ control respectively the position of dipole zero and the spatial decay of the dipole magnitude. A global multiplicative constant is absorbed into the light-matter coupling strength $\eta_{\rm c}$. This functional form is sufficiently flexible to describe both polar and non-polar molecules. For example, when $c_u=x_{\rm m}^{g,0}$, the permanent dipole moment vanishes at the neutral-state equilibrium geometry, corresponding to a non-polar molecule.

\subsubsection{Single photonic mode}
We begin by coupling the reactive molecule to a single cavity mode and analyzing its influence on the dissociative reaction dynamics.
\Fig{fig4:insidesidecavityratio1}~a) displays the ratio of the reaction rates inside and outside the cavity $k_{\rm c}/k_{\rm o}$ as a function of the cavity frequency $\omega_{\rm c}$, which we scan from 300 to $3500\,\mathrm{cm}^{-1}$ in increments of $10\,\mathrm{cm}^{-1}$. For this analysis, we consider a dipole function that vanishes asymptotically in the large $x_{\rm m}$ limit for a polar molecule: $u(x_{\rm m})=x_{\rm m}e^{-x_{\rm m}^2/\sigma_u^2}$ with $c_u=0$ and $a_{u}=2.5\,\text{\r A}$.  The light-matter coupling strength is fixed at $\eta_{\rm c}=0.1\,\mathrm{a.u.}$ and dissipation through the bosonic bath is omitted to isolate pure cavity effects. All following analyzes focus on the resonant transport regime at an applied bias of $\Phi=3{\rm V}$, where the molecular orbital lies well inside the bias window defined by the two chemical potentials at $-\Phi/2$ and $\Phi/2$. In this far-from-equilibirium regime, charge-driven excitation strongly accelerates the dissociation dynamics,  yielding a free-space (no-cavity) reaction rate of $k_{\rm o}=3.28\times 10^{-5}\,\mathrm{fs}^{-1}$. 

Unlike harmonic-oscillator eigenstates, the vibrational eigenstates of the Morse potential lack well-defined parity. Thus, all dipole-mediated transitions $\langle v_i| u(x_{\rm m}|v_{i'}\rangle$ are, in principle, allowed. In addition, the diagonal dipole matrix elements $\langle v_i|u(x_{\rm m})|v_i\rangle$ exhibits strong variation with the vibrational quantum number $v_i$. These features, combined with the adopted dipole function that decays asymptotically, give rise to the structure observed in \Fig{fig4:insidesidecavityratio1}~a): the rate modification reveals a series of sharp and well-resolved dips, i.e., resonant suppression peaks in $k_{\rm c}/k_{\rm o}$,  whenever the cavity frequency is tuned into close resonance with a vibrational transition. These include both nearest-neighbor transitions $v_i\leftrightarrow v_{i+1}$ and overtone transitions $v_i\leftrightarrow v_{i+2}$ among the bound states. The appearance of these discrete resonant-suppression features underscores the potential of employing a frequency-adjustable optical cavity to probe the intrinsic vibrational anharmonicity that governs reactive dynamics under non-equilibrium conditions. 

When the cavity frequency matches a particular vibrational transition, energy can be efficiently funneled into the cavity mode and subsequently dissipated into the cavity bath. This process creates an effective cooling pathway that selectively depletes population from high-lying vibrational levels, thereby suppressing access to the dissociation region and recuding the overall reaction rate. The mechanism is corroborated by the vibrational distribution dynamics  
$P_{v_i}(t)$. After rescaling by the surviving population $1-Q_{\rm loss}(t)$, the vibrational distribution evolves toward steady values at long times, with the final distribution strongly depends on the cavity frequency. As shown in \Fig{fig5:vibpop}, the steady populations 
$P_{v_i}(t)/(1-Q_{\rm loss}(t))$ for bound vibrational states below the dissociation barrier clearly illustrate this cavity-induced redistribution of vibrational population for two representative resonance conditions.
 
When $\omega_{\rm c}=2310\,\mathrm{cm}^{-1}$, resonant with the fundamental vibrational transition $v_0\leftrightarrow v_1$, we observe synchronized Rabi oscillations between these two states, emerging at approximately $200$~fs and damping out by around $6$~ps. The osillation period agrees precisely with $2\pi/\omega_c$. More importantly, the long-time populations exhibit a substantial enhancement of the ground-state population, a slight increase in the $v_1$ population, and a clear reduction in the $v_2$ population, compared to the cavity-free case (indicated by the colored circles aligned with the right axis). 

For $\omega_{\rm c}=2010\,\mathrm{cm}^{-1}$, resonant with the $v_1\leftrightarrow v_2$, the coherent oscillations shift accordingly to the $v_1$ and $v_2$ populations, with a slightly longer period reflecting the reduced $\omega_{\rm c}$. At long times, the populations of these two states increases, while the population of $v_3$ decreases. This pattern persists across frequencies: whenever $\omega_{\rm c}$ matches a specific transition $v_i\leftrightarrow v_{i+1}$, the intermediate-time dynamics display oscillations primarily in those two states, and the steady distribution shows enhanced populations in $v_i$ and $v_{i+1}$, accompanied by a corresponding depletion of the next higher level $v_{i+2}$. 

Further analysis of the dissociation probability resolved on different electronic states unveils a subtle asymmetry in how cavity frequency influences the reaction.
When dissociation proceeds on the charged surface, i.e., the molecule breaks apart as a charged species, stronger rate suppressions occurs for higher-frequency vibrational transitions low on the vibrational ladder. This trend becomes even more pronounced in an asymmetric molecule-lead coupling model, where under a positive bias the population resides almost entirely on the charged surface. In that limit, the only substantial rate suppression feature is the peak centered at the frequency corresponding to the fundamental transition $v_0\leftrightarrow v_1$, as presented in the supplementary material. Conversely, dissoication events originating from the neutral electronic state are most strongly inhibited when the cavity frequency is tuned closer to lower-frequency transitions higher up the ladder. These transitions involve more extended regions of the anharmonic potential and may play a more significant role in neutral-state bond cleavage. The divergence between the charged and neutral dissociation highlights that cavity-induced cooling does not act uniformly across electronic configurations. 

The stronger the light-matter coupling, the more effectively the cavity counteracts vibrational heating and inhibits dissociation. This is clearly demonstrated in \Fig{fig4:insidesidecavityratio1}~b), where the ratio $k_{\rm c}/k_{\rm o}$ is plotted against the light-matter coupling strength $\eta_{\rm c}$ for the cavity frequencies tuned near resonance with all nearest-neigbhor transitions among the bound vibrational levels. Interestingly, across the entire $\eta_{\rm c}$ range examined, the strongest overall rate suppression does not occur at the fundamental transition $v_{0}\leftrightarrow v_1$, even though these two states carry the largest population at room temperature. Instead, the maximum suppression appears midway up the vibrational ladder, at the transition $v_2\leftrightarrow v_3$. This behavior suggests that cavity-mediated cooling is not simply governed by thermal population factors.

In great contrast to the considerable rate modifications enabled by cavity-vibration resonance, the modulation of the molecular conductivity is found to be almost negligible. This is demonstrated in \Fig{fig6:insidesidecavitycurrent}, which displays the electronic current--rescaled by the surviving population, $I_e/(1-Q_{\rm loss})$ evaluated at $t=10\,\mathrm{ps}$, as a function of the cavity frequency. Although a faint resonant structure is discernible, the maximal current variation occurs at the photonic frequency $\omega_{c}=\Delta E_{v_0\leftrightarrow v_1}=2310\,\mathrm{cm}^{-1}$, and its magnitude is only about 0.2\%. This minimal effect is consistent with the extremely weak sensitivity of the electronic-state populations to the cavity frequency.  The cavity frequency lies in the infrared, far blow the charging energy, and the cavity mode does not participate in the molecule-electrode coupling. Thus, tuning the cavity in this case does not significantly influence the electronic-state transitions or the instantaneous changes of electronic populations that govern the electronic current. 

Nevertheless, a recent experimental work reported a dramatic electrical conductance enhancement by nearly six orders of magnitude when the cavity frequency is tuned to strongly couple an aromatic C-H out-of-plane bending vibration.\cite{Kumar_2024_JACS_p18999} Such modes are known to mediate diabatic couplings between electronic states and may therefore offer a direct and chemically grounded mechanism for modifying charge transport. Our prior theoretical study also demonstrated that a non-reactive out-of-plane bending vibration can serve as an efficient gateway that redirects population flow between different charged states, thereby opening an ultrafast reaction channel by facilitating diabatic hoppings from a bound potential energy surface to a repulsive one.\cite{Ke_2023_JCP_p24703} Extensions of the present simulations to include strong coupling between the cavity mode and a nuclear mode that either directly mediates the diabtatic coupling between distinct charged states or modulates the molecule-electrode contact geometry (e.g., a tethering bond linking the molecule to the electrodes) is an important direction for future work. Incorporating such a mode would allow us to gain deeper insights into how vibrational polaritons influence molecular conductivity and dissociative behavior in the strong coupling regime. 

\begin{figure}
\centering
  \begin{minipage}[c]{0.5\textwidth}
    \includegraphics[width=\textwidth]{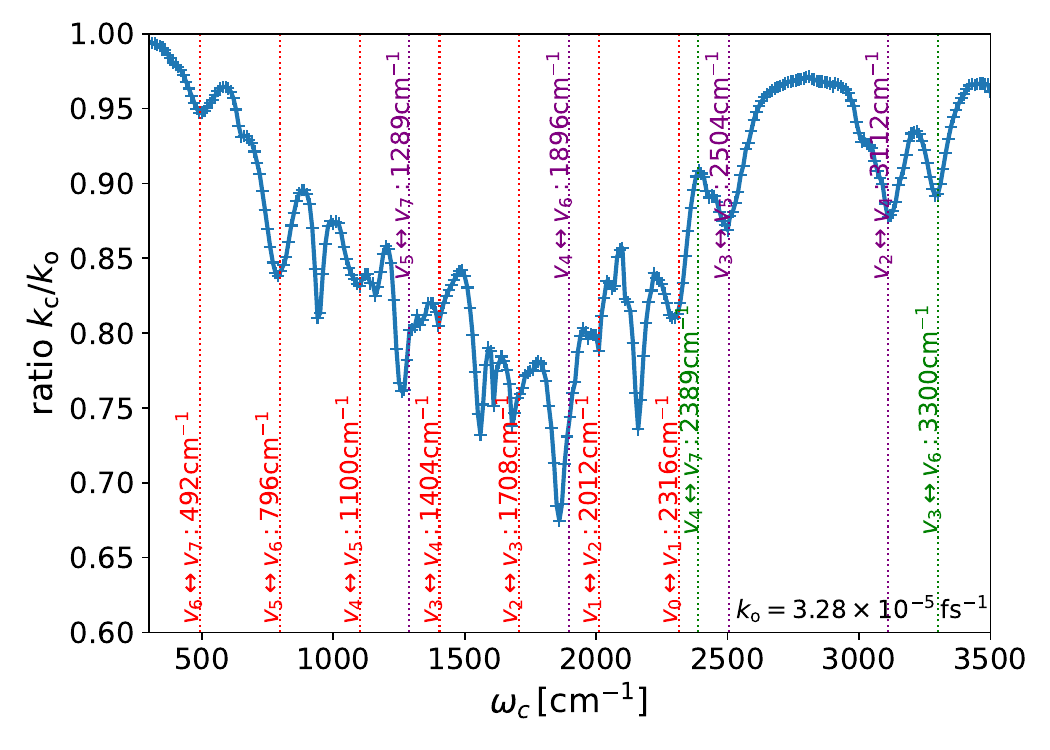}
 \end{minipage}
\caption{Same as \Fig{fig4:insidesidecavityratio1}~a), except that the electric dipole function is non-polar and adopts an undamped form, $u(x_{\rm m})=x_{\rm m}-x_{\rm m}^{g,0}$.} \label{fig7:insidesidecavityratio2}
\end{figure}

\begin{figure}
\centering
  \begin{minipage}[c]{0.5\textwidth}
    \includegraphics[width=\textwidth]{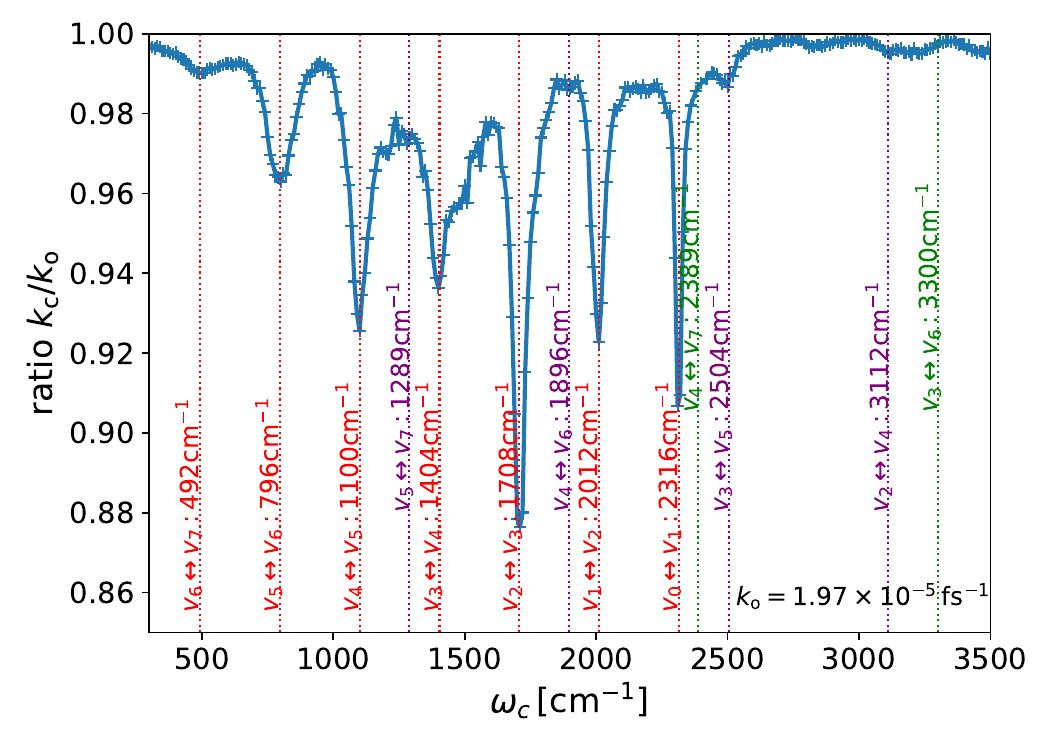}
 \end{minipage}
\caption{Same as \Fig{fig4:insidesidecavityratio1}~a), except that the dissipation strength to the bosonic bath is set to $\lambda_{\rm m}=50\,\mathrm{cm}^{-1}$.} \label{fig8:insidesidecavityratio3}
\end{figure}

Since cavity-induced vibrational transitions are strongly governed by the molecular dipole function, in \Fig{fig7:insidesidecavityratio2}, we present results
obtained using a different electronic dipole function, $u(x_{\rm m})=x_{\rm m}-x_{\rm m}^{g,0}$, which does not decay in the $x_{\rm m}\rightarrow \infty$ limit. In this case, the cavity field couples to large-amplitude molecular motion along the reactive coordinate. For the same light-matter coupling strength, this choice leads to a noticably stronger overall rate suppression effect. Compared with \Fig{fig4:insidesidecavityratio1}~a), the suppression profile also becomes broader, and multi-quantum vibrational cooling is more prominently favored by the cavity. The strongest peak now corresponds to the overtone transition $v_6\rightarrow v_4$, and transitions involving more than two vibrational quanta, for example, $v_6\rightarrow v_3$ are also activated. Furthermore, several peaks appear at $\omega_{\rm c}=940\,\mathrm{cm}^{-1}$, $1550\,\mathrm{cm}^{-1}$,  $2160\,\mathrm{cm}^{-1}$ that cannot be assigned to bound-bound transitions and likely originate from transitions connecting a bound vibrational state to the continuum region beyond the dissociation barrier. Additional results and discussion for other choices of $u(x_{\rm m})$ are provided in the supplementary material. 

Our conclusions remain qualitatively robust across different choices of system and bath parameters. Specifically, the characteristic multi-peak resonant profile in the cavity-modified reaction rates is consistently preserved. 
However, the relative intensities of the individual rate suppression peak can be subtly influenced by these parameters. For example, \Fig{fig8:insidesidecavityratio3} shows results obtained with a finite value of $\lambda_{\rm}=50\,\mathrm{cm}^{-1}$, which accounts for excitation quenching caused by dissipation into the solvent or surface phonons. Compared with \Fig{fig4:insidesidecavityratio1}~a) for $\lambda_{\rm}=0$, the resonant structure remains similar but exhibits slightly altered peak strengths.  Other factors, such as the vibronic coupling strength (proportional to $x_{\rm e}^{e,0}-x_{\rm m}^{g, 0}$, the applied bias voltage $\Phi$, the molecule-electrode coupling configuration and bias polarity, and the cavity damping rate $\lambda_{\rm c}$, also modulate the detailed shape of the rate modification profile in subtle ways. Additional results presenting variations with different $x_{\rm e}^{e,0}$, $\Phi$,  $\lambda_{\rm c}$, and asymmetric molecule-electronic coupling scenarios are provided in the supplementary material.

Finally, we note a practical implication for future experimental studies: in realistic molecular junctions, a chemical reaction such as bond rupture or configuration switching often manifests itself as abrupt changes in electrical conductivity. The delay time at a fixed applied bias voltage before a measurable conductivity shift occurs may thus provide an experimentally accessible proxy for the underlying chemical reaction rate. Based on our findings, a cavity-induced slowdown of bond rupture should translate directly into a correspondingly longer delay before the conductance drop, offering an alternative and reliable means of quantifying cavity-induced reaction kinetics. More broadly, driving the molecule out of equilibrium allows access to strongly anharmonic regions of the potential surface, potentially broadening the scope of chemical reactions that can be explored and controlled under vibrational strong coupling in a confined electromagnetic environment. 

\begin{figure*}
\centering
  \begin{minipage}[c]{0.45\textwidth}
    \raggedright a) $u(x_{\rm m})=x_{\rm m}e^{-x^2_{\rm m}/a_{u}^2}$
    \includegraphics[width=\textwidth]{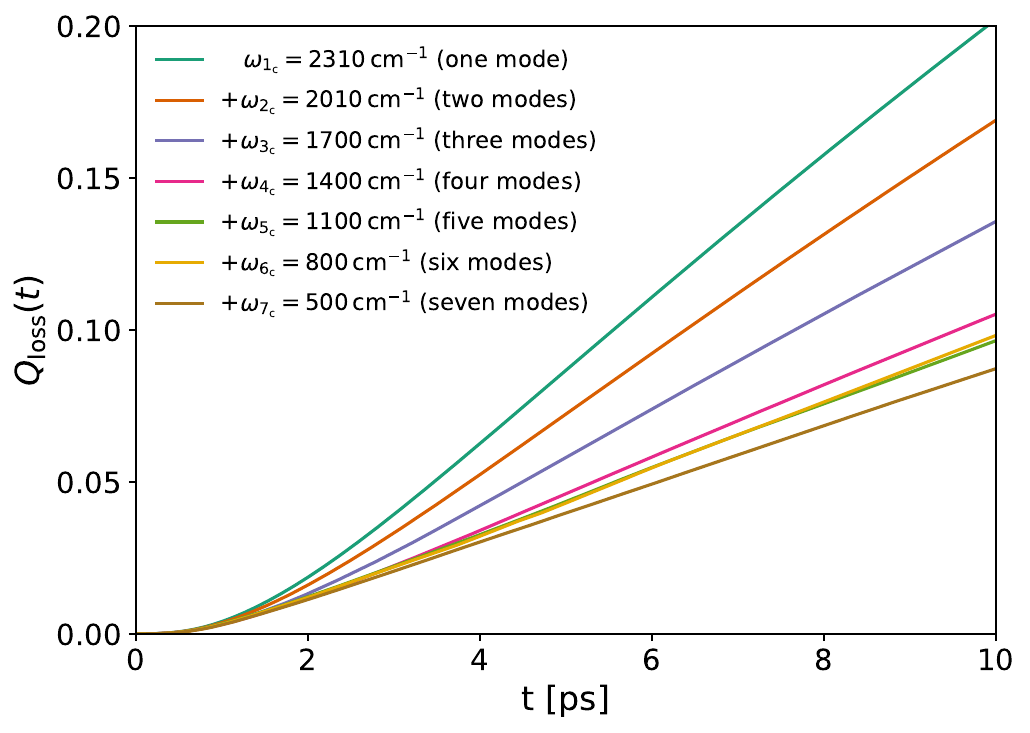}
 \end{minipage}
   \begin{minipage}[c]{0.4\textwidth}
    \includegraphics[width=\textwidth]{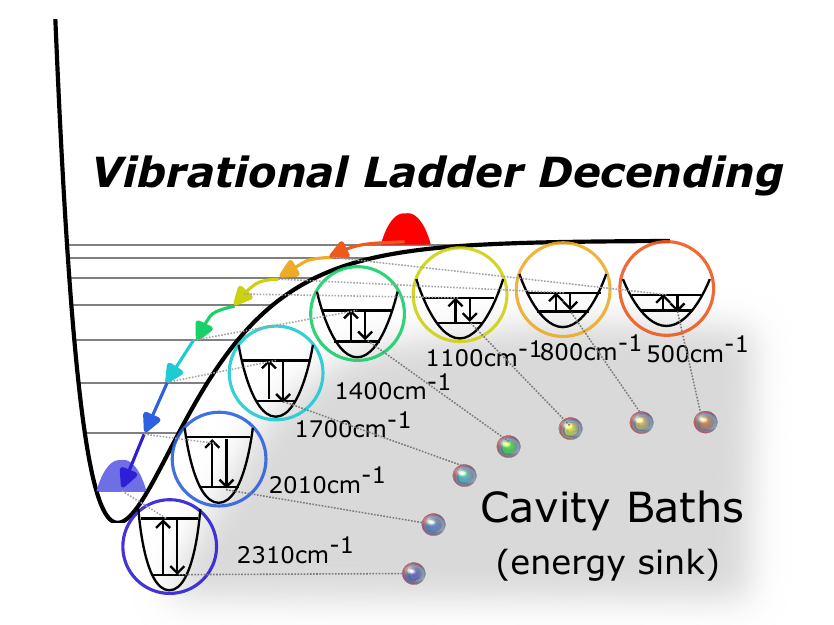}
 \end{minipage}
  \begin{minipage}[c]{0.45\textwidth}
    \raggedright b) $u(x_{\rm m})=x_{\rm m}-x_{\rm m}^{g,0}$
    \includegraphics[width=\textwidth]{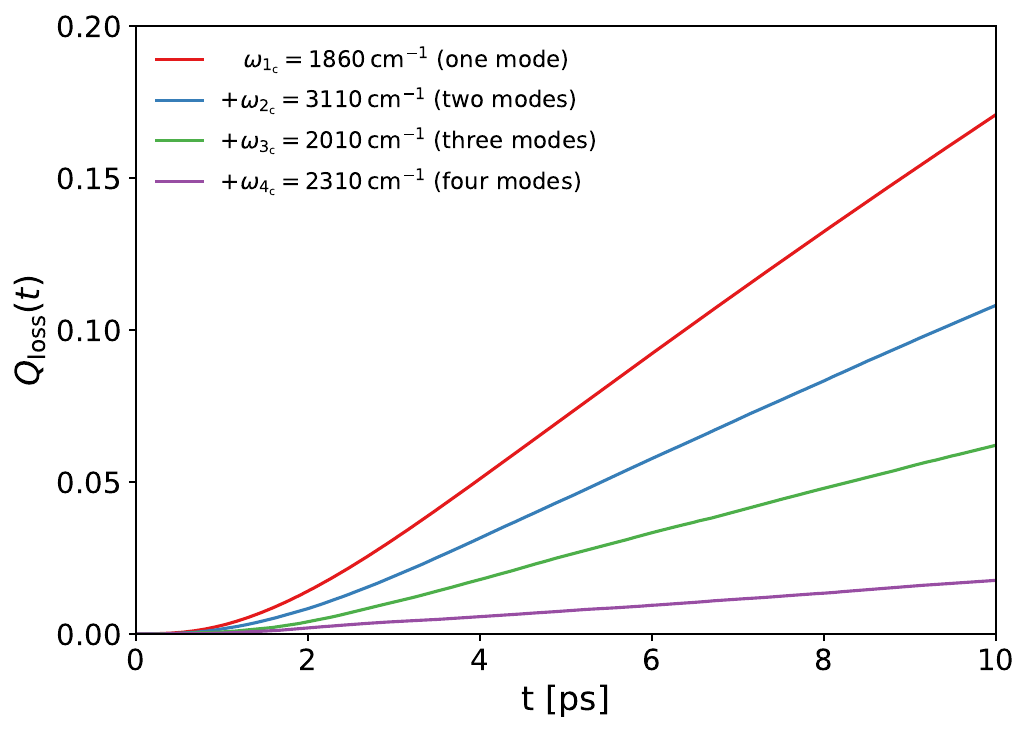}
 \end{minipage}
    \begin{minipage}[c]{0.4\textwidth}
    \includegraphics[width=\textwidth]{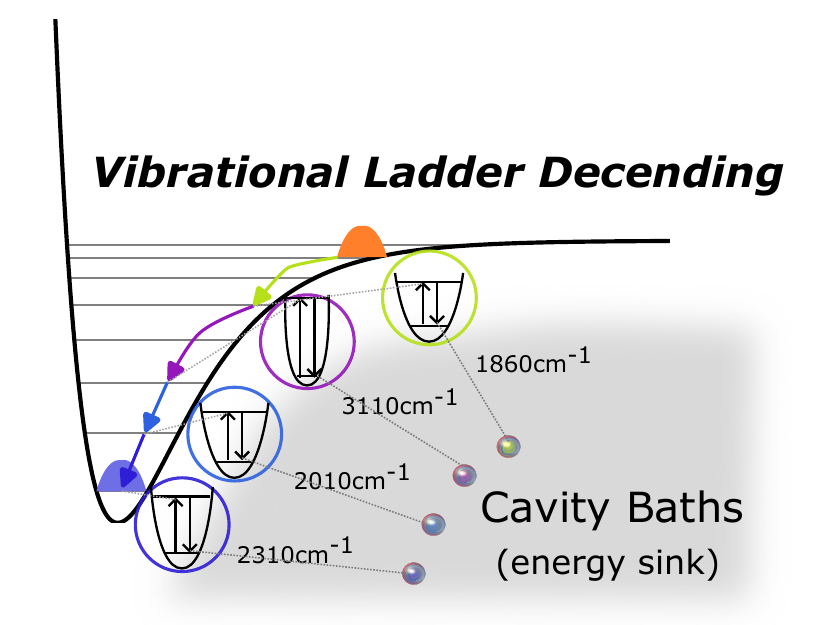}
 \end{minipage}
\caption{Time-dependent dissociation probability $Q_{\rm diss}(t)$ in an $N_{\rm c}$-mode cavity where $N_{\rm c}\ge1$. a) Results obtained using the electronic dipole function $u(x_{\rm m})=x_{\rm m}e^{-x^2_{\rm m}/a_{u}^2}$ with $a_{u}=2.5\,\text{\r{A}}$. The number of caivty modes $N_{\rm c}$ is increased from 1 to 7 with each additional cavity mode tuned into near resonance with a vibrational transition between adjacent levels along the anharmonic vibrational ladder below the dissociation barrier. The resulting multi-mode strong coupling opens up an increasingly effective dissipation channel via a stepwise vibrational ladder descending process, illustrated schematically on the right. b) Same as panel a), but using the undamped dipole function $u(x_{\rm m})=x_{\rm m}-x_{\rm m}^{g,0}$, which more efficiently promotes multi-quantum vibrational cooling. In this case, a four-step relaxation channel $v_6\rightarrow v_4\rightarrow v_2\rightarrow v_1\rightarrow v_0$ is activated,  effectively transferring the vibrational population from near the dissociation threshold down to the potential bottom in the vibrational ground state, as shown schematically on the right.  In both panels, the bias voltage is fixed at $\Phi=3$~V, with light-matter coupling $\eta_{\rm c}=0.2$~a.u. and no molecular dissipation to the bosonic bath $(\lambda_{\rm m}=0)$. } \label{fig9:insidesidecavityLoss}
\end{figure*}

\begin{figure}
\centering
  \begin{minipage}[c]{0.5\textwidth} 
    \includegraphics[width=\textwidth]{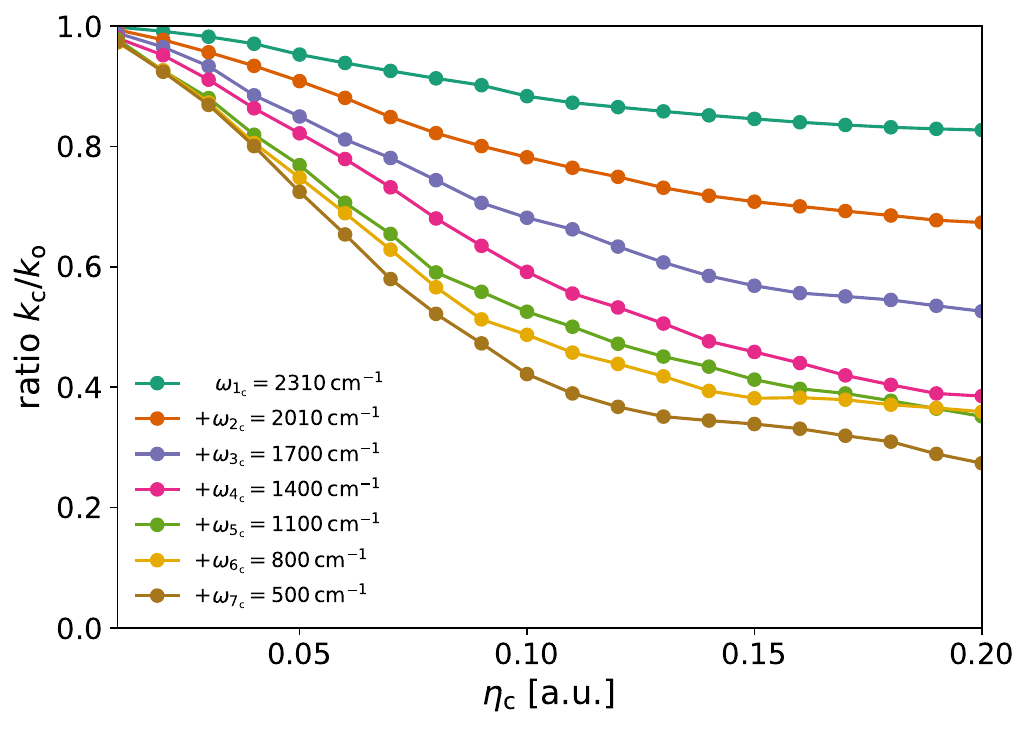}
 \end{minipage}
\caption{Ratio of reaction rates $k_{\mathrm c}/k_{\mathrm{o}}$ in multi-mode cavities as a function of the light-matter coupling strength $\eta_c$, which is taken to be identical for all cavity modes. The electronic dipole function is $u(x_{\rm m})=x_{\rm m}e^{-x^2_{\rm m}/a_{u}^2}$ with $a_{u}=2.5\,\text{\r{A}}$. The bias volgate is fixed at $\Phi=3$~V, and the molecular dissipation to the bosonic bath is neglected, i.e., $\lambda_{\rm m}=0$. } \label{fig10:insidesidecavityratio}
\end{figure}

\subsubsection{Multiple photonic modes}
A major challenge in fabricating stable molecular electronic devices is the poor robustness of single-molecule junctions, particularly under high bias voltages.\cite{Stipe_1997_PRL_p4410,Persson_1997_SS_p45,Kim_2002_PRL_p126104,Huang_NanoLett._2006_p1240,Kuznetsov_Electrochem.Commun._2007_p1624,Schulze_Phys.Rev.Lett._2008_p136801,Ioffe_Nat.Nanotechnol._2008_p727,Sabater_2015_BJN_p2338,Li_2016_JACS_p16159,Peiris_Chem.Sci._2020_p5246,Bi_J.Am.Chem.Soc._2020_p3384}  In a current-carrying molecular junction, strong vibronic coupling within the central organic molecule efficiently funnels the excess energy carried by injected charge carriers into critical vibrational modes. This rapid vibrational heating often outpaces energy dissipation into the electrodes and surrounding environment, and ultimately trigger destructive chemical bond cleavage. 

As shown in the previous section, embedding the junction inside a confined single-mode optical cavity provides an effective, and more importantly, selective vibrational energy relaxation channel. By hybridizing a targeted molecular vibrational transition with a cavity mode, a photon-assisted relaxation pathway opens, hindering a step of voltage-induced heating and reducing population flow further into high-energy vibrational states that are precursors to chemical degradation. Motivated by the inherent anharmonicity of reative molecular potentials, without which chemical reactions would not occur, we extend this idea by proposing a cascading vibrational ladder descending mechanism. In this scheme, a sequence of cavity modes is strategically tuned to match the anharmonic spacings of vibrational levels. Such a multi-mode vibrational strong coupling architecture could, in principle, enable multiple resonant energy-extraction steps, each selectively draining energy from a distinct rung of the vibrational ladder.  

To assess the feasibility of this concept, we examined the dissociation dynamics while incrementally adding cavity modes, each tuned into near resonance with a specific vibrational transition along the anharmonic vibrational ladder. The time-evolving dissociation probabilities are shown in \Fig{fig9:insidesidecavityLoss} for two representative dipole functions. In these simulations, the coupling strength to each photonic mode is assumed to be identical and fixed at $\eta_{\rm c}=0.2$~a.u. In line with our expectations, the inclusion of additional cavity modes, whether resonant with nearest-neighbor vibrational transitions or with overtone transitions, systematically slows the growth of $Q_{\rm loss}(t)$, demonstrating a substantial suppression of dissociation. 

This multi-mode strong coupling strategy facilitates rate deceleration by enabling each cavity mode to siphon off energy from a different step of the vibrational heating sequence, thereby pulling the wavepacket downward and away from the dissociation threshold more effectively. This mechanism is schematically illustrated on the right side of  \Fig{fig9:insidesidecavityLoss}. Notably,
the engineered vibrational ladder descending process proposed here is conceptually opposite to the well-estabilished vibrational ladder climbing, which has been realized in optimal control experiments using negatively chirped strong infrared laser pulses, \cite{Chelkowski_1990_PRL_p2355,Witte_2003_JCP_p2021,Marcus_2006_EL_p43} or in tunneling junctions.\cite{Tikhodeev_2005_SS_p25,Nacci_2009_NL_p2996} Here, however, the energy extraction is achieved not via external high-intensity laser pulses but through the fluctuating photons of the confined cavity field. 

The resulting rate suppression in an $N_{\rm c}$-mode cavity, with $N_{\rm c}$ ranging from 1 to 7 and using the dipole function $u(x_{\rm m})==x_{\rm m}e^{-x^2_{\rm m}/a_{u}^2}$, is shown in \Fig{fig10:insidesidecavityratio} as a function of the light-matter coupling strength $\eta_{\rm c}$. Each cavity mode is tuned sequentially to closely match a distinct vibrational transition with a progressively smaller energy gap along the vibrational manifold.  As expected, stronger coupling yields more pronounced cooling. For $N_{\rm c}=7$ and $\eta_{\rm c}=0.2$~a.u., the reaction rate is reduced to approximately one-fifth of its cavity-free value. 

The system undergoes a sequence of vibrational heating and cavity-assisted cooling processes along the vibrational ladder, which may not be apparent in linear spectroscopy. Looking ahead, a promising extension of this work would be to simulate surface-enhanced ultrafast 2D-IR spectra,\cite{Kraack_2016_PCCP_p16088,Kraack_2016_JPCC_p3350,Kraack_2017_CR_p10623} which have proven effective at capturing nonlinear and high-order vibrational signals. Such simulations\cite{F.Ribeiro_2018_JPCL_p3766,Li_2025_PRR_p33248,Hirschmann_2024_N_p2029} could provide a more intuitive, real-time visualization of the non-equilibrium vibrational energy flow described here, paving the way for a broad range of future investigations that combine nanostructured surfaces, confined optical fields, and femtosecond IR spectroscopy to explore the ultrafast dynamics of highly excited vibrational states, energy transfer pathways, and coherent control at surfaces.\cite{Yang_2023_JPCB_p2083,Gilbert_2025_JACS_p40099}

\section{\label{sec:conclusion}Conclusion}
In this work, we have presented a theoretical investigation of current-driven non-equilibrium chemical reactions occurring at a molecule-electrode interface embedded in a frequency-tunable, confined electromagnetic vacuum cavity. 
By driving the molecular system far from thermal equilibrium, one gains access to reaction pathways and highly excited vibrational states that are otherwise inaccessible under ambient thermal conditions. As a consequence, this non-equilibrium control may open new opportunities for probing intrinsic molecular anharmonicity and expanding the scope of polaritonic chemistry beyond the conventional near-equilibrium vibrational strong coupling regime. 

To rigorously investigate these phenomena, our simulations were conducted within a fully quantum-mechanical framework, combining the open quantum dynamical approach--HEOM method--with a tree tensor network state solver. This methodology provides a numerically exact treatment of the strong and intricate correlations among photonic, molecular electronic, and vibrational DoFs, while fully accounting for their interactions with electronic reservoirs and dissipative bosonic baths.

The results reveal a remarkable resonant rate suppression when an infrared cavity mode is tuned into near resonance with either a nearest-neighbor or overtone vibrational transition along the reaction coordinate. This observation highlights the potential of current-carrying molecular optoeletronic junctions as sensitive probes of molecular intrinsic anharmonicity, a defining ingredient of chemical reactivity. Building on our single-mode analysis, we further proposed a cavity-engineered vibrational cooling scheme in which multiple cavity modes are strategically tailored to match distinct vibrational transition energies in accordance with the reactive anharmonicity. Such a multi-mode strong coupling design enables a cascading vibrational ladder descending process, whereby vibrational energy is successively drained from higher to lower excited states. This cooling mechanism may offer a viable route to suppress voltage-induced bond rupture and thereby mitigate the long-standing stability challenges of molecular junctions under high bias.

At present, the work presented here remains a hypothetical, proof-of-concept study that, to the best of our knowledge, has not yet been realized in any laboratorical experiment. While we hope that the ideas introduced here will inspire future experimental efforts, we are also cognizant of the oversimplifications inherent in our model. Future extensions of this work will involve exploring more complex systems and scenarios, incorporating additional internal DoFs that yield a more realistic description of the reactive species, as well as photonic modes in the visible range that can directly promote electronic transitions.  Furthermore, whereas we have assumed identical temperatures for the electrodes, introducing a thermal gradient across two terminals bridged by the molecule represents another important direction for future investigation. Such studies may provide deeper insight into how the vacuum electromagnetic field modulates heat and energy transport in molecules and nonequilibrium molecular devices.

\begin{acknowledgments}
The author thanks the Swiss National Science Foundation for the award of a research fellowship (Grant No. TMPFP2\_224947). 
\end{acknowledgments}

\section*{Supplementary information}
See the supplementary material for: 1) Additional analysis of the nonequilibrium reaction and transport dynamics outside the cavity; 2) Further results illustrating the influence of the electrionic dipole function, bias voltage, vibronic coupling strength, and cavity loss rate on cavity-induced rate modifications under nonequilibrium conditions; 3) An extension to scenarios involving asymmetric molecule-electrode coupling. 

\section*{Data Availability Statement}
The data and code that support the findings of this work are available from the corresponding author upon reasonable request.

\end{document}